\documentclass[12pt]{article}

\usepackage[numbers, sort&compress]{natbib}
\usepackage{fullpage}
\usepackage{epsfig}
\usepackage{latexsym,amsmath,amssymb,amsthm}
\usepackage{subfig}
\newcommand{\ceil}[1]{\left\lceil {#1} \right\rceil}

\def\maps11{\stackrel {1-1}{\longmapsto}}

\begin{document}

% \title{Coupled IEEE 802.11ac and TCP Goodput improvement
% using Aggregation and Reverse Direction}

% \title{Optimizing TCP Goodput and Delay in next generation IEEE 802.11 (ax) devices}

\title{Advanced IEEE 802.11ax TCP aware scheduling under unreliable channels}

\author{%
Oran Sharon
\thanks{Corresponding author: oran@netanya.ac.il, Tel: 972-4-9831406,
Fax: 972-4-9930525} \\
Department of Computer Science \\
Netanya Academic College \\
1 University St. \\
Netanya, 42365 Israel
\and
Yaron Alpert\\
Intel\\
13 Zarchin St.\\
Ra'anana, 43662, Israel\\
Yaron.alpert@intel.com
}

% \fi %%%%%

\date{}

\maketitle

\begin{abstract} 
In this paper we suggest 
advanced IEEE 802.11ax TCP-aware scheduling 
strategies for optimizing the AP operation
under transmission of unidirectional TCP traffic.
Our scheduling strategies optimize the
performance using
the capability for Multi User transmissions 
over the Uplink, first introduced in IEEE 802.11ax,
together with Multi User transmissions over the Downlink.
They are based on Transmission Opportunities (TXOP) 
and we suggest three scheduling
strategies determining the TXOP formation parameters.
In one of the strategies one can control the achieved 
Goodput vs. the delay. We also assume
saturated WiFi transmission queues. We show
that with minimal Goodput degradation one can avoid
considerable delays.
% one can achieve over 90 percent of the maximum Goodput
% in cycles of 3-5 {\it ms} compared to cycles of 25 {\it ms}
% which achieve the maximum Goodput.
\end{abstract}

\bigskip


\renewcommand{\baselinestretch}{1.3}
\small\normalsize

%%%%%%%%%%%%%%%%%%%%%%%%%%%%%%%%%%%%%%%%%%%%%%%%%%%%%%%%%%%%%%%%

\section{Introduction}

\subsection{Background}

In 2013 the IEEE established a new Task Group (TG),
denoted TGax, in order to establish a new standard
for the IEEE 802.11 networks~\cite{IEEEBase1}. The functional
requirements for IEEE 802.11ax, also denoted High
Efficiency WLAN (HEW)~\cite{IEEEax} are to achieve at least four
times as much improvement in the average throughput per station
compared to former versions of the IEEE 802.11 standards~\cite{IEEEac},
and to support dense deployment environments~\cite{PS,KKL,AVA,DCC,B}.

In order to achieve these goals, several new features are
considered by TGax including physical layer techniques,
medium access control layer strategies, spatial frequency
reuse schemes and power saving mechanisms.

In the physical layer the new techniques are adopting
Orthogonal Frequency Division Multiple Access (OFDMA)
and deploying Multi User Multiplex-Input
Multiplex-Output (MU-MIMO) over the DL and UL.

Concerning the MAC layer, in order to improve the Basic Service
Set (BSS) throughput, it is crucial to develop an efficient
MAC scheme which can reduce the probability of a transmission
collision among different stations, allow for simultaneous
transmissions in the same BSS  and decrease the channel
time for transmission of control information. One way
to achieve the above goals is to improve the basic
access mechanism in IEEE 802.11 networks, namely
Distributed Coordination Function (DCF) based
on Carrier Sense Multiple Access with Collision
Avoidance (CSMA/CA). The second way is to
focus on new MAC schemes for simultaneous multi user
transmissions, based on the OFDMA and MIMO technologies.

\subsection{Research question}

One of the remarkable additions in IEEE 802.11ax is to enable
MU-MIMO and/or OFDMA
over the Uplink, with combination of MU-MIMO and/or OFDMA
over the Downlink, where only DL MU-MIMO was
possible in the predecessor of IEEE 802.11ax,
namely IEEE 802.11ac~\cite{IEEEac}.
MU over the UL enables simultaneous transmissions
of several stations to the AP. In this paper we evaluate
several scheduling strategies in order to optimize 
the performance of MU over the Downlink and Uplink in
relation to TCP traffic which is composed of two way traffic;
one way for the TCP Data segments and the other way for the
TCP Ack segments.

The current paper is a 
continuation of papers~\cite{SA1,SA2,SA3,BS,SA4}.
In papers~\cite{SA1,SA2,SA3,BS} the authors suggest
scheduling strategies for the parallel transmissions of the AP to
a given set of stations using new features
of IEEE 802.11ax . The authors assume UDP-like
traffic where the AP transmits data MAC Service Data
Units (MSDUs) to the stations,
which reply with MAC acknowledgments. 
In paper~\cite{SA4} and in this paper  we
assume a DL unidirectional TCP-like traffic in which the AP transmits
TCP Data MSDUs to a given set of stations, and the stations reply
with TCP Ack MSDUs. 
The differences between this paper and paper~\cite{SA4}
is that in paper~\cite{SA4} the assumption is that
the channel is reliable, while in this paper we extend
the work to unreliable channels, which opens the
door for the use of advanced transmission strategies and 
considers the impact of various channels as reflected by
the use of different MCSs over the DL and UL.
As far as we know
the issue of transmitting
TCP traffic over IEEE 802.11ax has not yet been investigated,
except in paper~\cite{SA4}. 
We suggest several TCP-aware scheduling
strategies for transmission of 
unidirectional TCP Data segments over
the DL and TCP Ack segments over the UL using
Multi User (MU) mode
for 4, 8, 16 and 32
stations scenarios over an unreliable
channel. 
This is one of the aspects to compare between new
amendments of the IEEE 802.11 standard~\cite{KCC}.
In this paper we are interested in finding a theoretical 
upper bound on the maximum 
DL unidirectional TCP Goodput
that can be achieved by IEEE 802.11ax without any
delay restriction and 
comparing between the various TCP-aware scheduling strategies.
Therefore, due to a very short round trip delay we assume
the traffic saturation model
where TCP connections always have data to transmit
( no limit on the amount of TCP Data segments waiting
and on the TCP transmission queue )
and the TCP Ack is generated immediately by receivers.
Second, we neutralize any aspects
of the PHY layer as the
number of Spatial Streams (SS) in use and channel correlation
when using MU, the use in the sounding protocol etc.
We also use equal Resource Units (RUs) allocation 
for all served stations and optimize the MU-MIMO/OFDMA in order
to reduce the overhead and to increase the Goodput.

\indent
As mentioned, we assume that every TCP connection has
enough TCP Data segments to fulfill the next PPDU
and we assume that
transmissions are made using an optimized
( in terms of overhead reduction and Goodput 
maximization ) two level aggregation scheme
to be described later.
Our goal
is to find an upper bound on
the maximum possible Goodput that the
wireless channel enables the TCP connections,
where the TCP itself does not impose any limitations
on the offered load, i.e. on the rate that MSDUs are
given for transmission to the
MAC layer of the IEEE 802.11ax.
We then optimize the delay within minimal
Goodput maximization.
We also assume that the AP and the stations
are the end points of the TCP connections.
Following e.g.~\cite{MKA,BCG1,BCG2,KAMG}
it is quite common to consider short Round Trip Times (RTT)
in this kind of high speed network such that
no retransmission timeouts occur. 
Moreover, we assume that every TCP connections' Transmission Window
can always provide as many MSDUs to transmit as the IEEE 802.11ax
protocol limits
enable. This assumption follows the observation
that aggregation is useful in a scenario where the offered
load on the channel is high. 

This research is an additional step in investigating TCP traffic
in IEEE 802.11ax. In our further papers we plan
to address other 
TCP traffic scenarios to investigate
such
as UL unidirectional TCP traffic and bi-directional TCP traffic
and to reflect other protocol limitations.

\subsection{Previous works}

Most of the research papers on IEEE 802.11ax
thus far examine new mechanisms in the PHY and MAC
layers of the IEEE 802.11ax proposal~\cite{C,OA,NY,QLYY,DLLC,AK,PH,LLYQYZY, LDC, KBPSL, JS, RFBBO, RBFB, HYSG,KLL}.
The issue of TCP traffic
over IEEE 802.11ac networks 
has been investigated, e.g.
in~\cite{SA10,SA11,SA12}, for uni-directional
DL TCP traffic, uni-directional UL TCP traffic
and both bi-directional DL and UL TCP traffic.
However, in all these works there is no possibility
of using the MU operation mode over the UL, a feature
that was first introduced in IEEE 802.11ax .
To the best of our knowledge, paper~\cite{SA4} is the first
and only  paper so far that deals with TCP traffic over
IEEE 802.11ax networks. In this paper the authors assume
a reliable channel and saturated TCP traffic and show
two-level aggregation efficient packing policy of
TCP Data/Ack segments in MAC Protocol Data Units (MPDUs).
They suggest three scheduling strategies for the
transmission of DL TCP Data segments based on  Single User (SU)
and OFDMA+MU-MIMO PHY layers, and consider 4, 8, 16, 32 and 
64 stations in the system and measure the TCP Goodput of the
system using TXOPs as will be described later
in this paper.
The main findings are that for 4 and 8 stations the MU strategies 
outperform the SU one, for 16 and 32 stations the MU and SU strategies
have about the same performance and in the case of 64
stations the SU strategy outperforms the MU ones. 

In this paper we extend previous works
by assuming an unreliable channel and so the
efficient packing used in~\cite{SA4} cannot be used. Therefore, 
we suggest another TCP-aware scheduling
strategy; consider the use of
different Modulation/Coding schemes (MCSs) over the DL/UL
which together with the bandwidth and
the Signal-to-Noise Ratio (SNR) influence
the Bit Error Rate (BER).

\indent
The remainder of the paper is organized as follows: 
In Section 2 we describe
the new mechanisms of IEEE 802.11ax 
relevant to this paper. In Section 3 we 
describe the operation mode and the TCP-aware
scheduling strategies
suggested.
We assume the reader is familiar 
with the basics of PHY and MAC layers
of IEEE 802.11 described in previous papers, e.g.~\cite{SA}. 
In Section 4 we present the simulation
results for the Goodput and Delay
of the various proposed TCP-aware scheduling strategies and
Section 5 summarizes 
the paper. 
In the Appendix we show an example
of the relation between the channel bandwidth,
Modulation/Coding schemes in use and
the Signal-to-Noise-Ratio to the received 
Bit Error Rate.
Lastly, moving forward, we denote IEEE 802.11ax by 11ax .

\section{New features in IEEE 802.11ax}

As mentioned, 11ax focuses on implementing 
mechanisms to efficiently serve more
users, enabling consistent and 
reliable streams of data ( average throughput
per user ) in the presence of multiple users. 
In order to meet these targets we mention a few
new 11ax mechanisms in both the PHY and
MAC layers, relevant to this paper. At the PHY layer, 
11ax enables larger OFDM FFT sizes (4X larger)
and therefore every OFDM symbol is 
$12.8 \mu s$ compared to $3.2 \mu s$ in 
IEEE 802.11ac, the predecessor of 11ax .
By narrower sub-carrier spacing (4X closer)
the protocol efficiency is increased because
the same Guard Interval (GI) is used both in 11ax and 
in previous versions of the standard.

In addition, to increase the average 
throughput per user in high-density scenarios,
11ax introduces two new Modulation Coding Schemes 
(MCSs), MCS10 (1024 QAM ) and MCS 11 (1024 QAM 5/6), 
applicable for transmission with bandwidth larger than 20 MHz.

In this paper we use the Transmission Opportunity (TXOP)
feature first introduced in IEEE 802.11n~\cite{IEEEn}.
This feature allows a station, after gaining
access to the channel, to transmit
several PHY Protocol Data Units
(PPDUs) in a row without interruption.
For scenarios
with bidirectional traffic such as TCP Data/Ack segments,
this approach is very efficient as it reduces contention
in the wireless channel and reduces delay by sending
the TCP Acks in the same TXOP.

We focus on optimizing 
the TXOP duration and pattern, PPDU duration and
the 11ax's two-level aggregation
scheme working point first introduced 
in IEEE 802.11n~\cite{IEEEn}, 
in which several  MPDUs can be aggregated 
to be transmitted in a  single PHY Service 
Data Unit (PSDU). Such aggregated PSDU 
is denoted Aggregate MAC Protocol Data
Unit (A-MPDU) frame. In two-level aggregation 
every MPDU can contain several 
MSDUs. 
MPDUs are separated by an MPDU 
Delimiter field of 4 bytes and 
each MPDU contains MAC Header
and Frame Control Sequence (FCS) fields.
MSDUs within an MPDU are separated 
by a SubHeader field of 14 bytes. 
Every MSDU is rounded to an 
integral multiple of 4 bytes 
together with the SubHeader field. 
Every MPDU is also rounded to 
an integral multiple of 4 bytes.

\begin{figure}
\vskip 9cm
\includegraphics{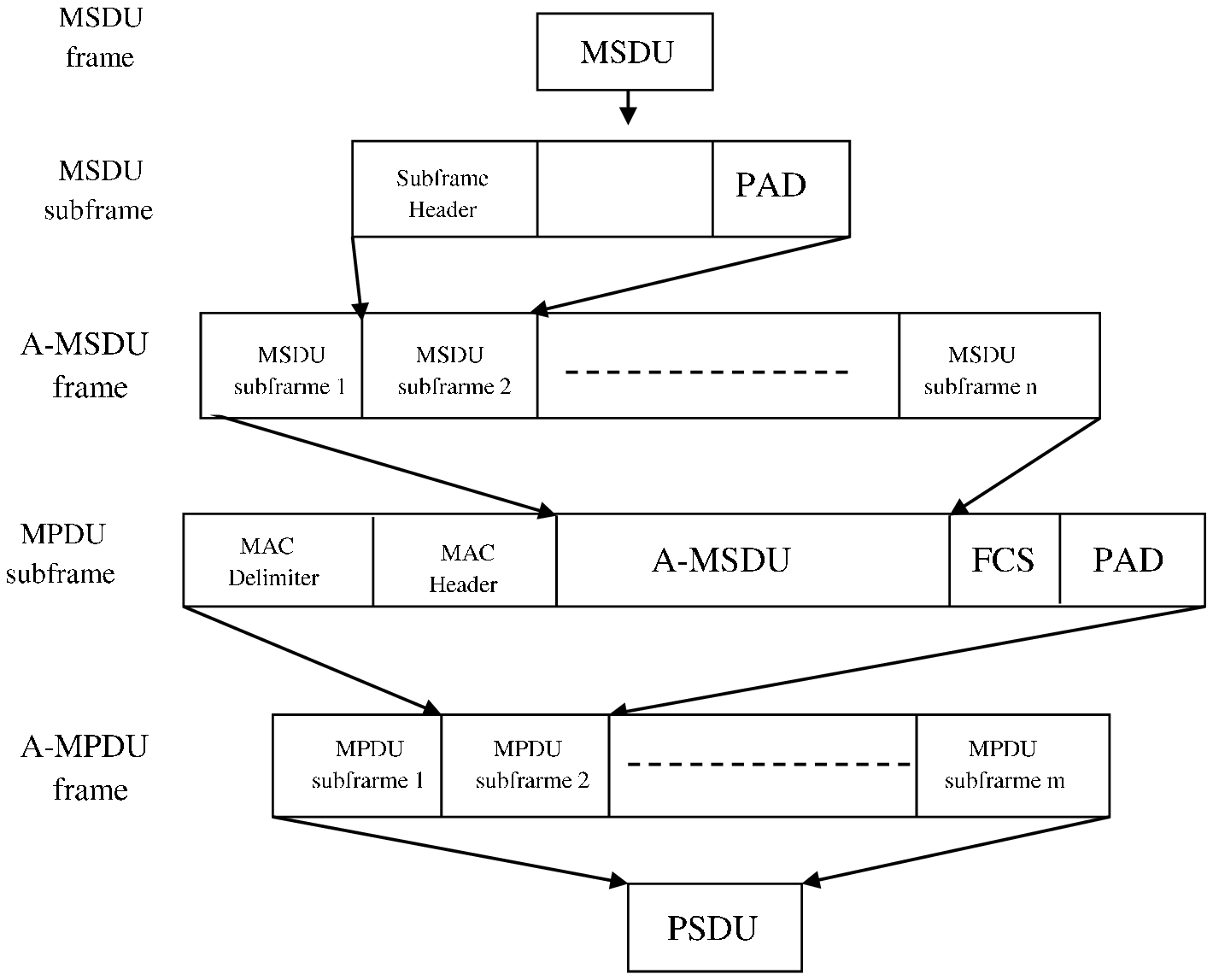}
\caption{The generation of an A-MPDU frame
in two-level aggregation.}
\label{fig:twole}
\end{figure}

In 11ax the size of 
an MPDU is limited to 11454 bytes and the
size of the A-MPDU frame is 
limited to 4,194,304 bytes.
The transmission time of the PPDU 
(PSDU and its preamble) is limited to 
$5.484ms$ ($5484 \mu s$)
due to the L-SIG (one of the legacy 
preamble's fields) duration limit~\cite{IEEEBase1}.
The A-MPDU frame structure in two-level aggregation
is shown in Figure~\ref{fig:twole}.

IEEE 802.11ax also enables extension of the MAC acknowledgment 
mechanism by using
an acknowledgment window of 256 MPDUs.
In this paper 
we also assume that all MPDUs 
transmitted in an A-MPDU frame are 
from the same Traffic Stream (TS). 
In this case up to 256 MPDUs are
allowed in an A-MPDU frame of 11ax. 

Recall that in 11ax it is also possible to use MU-MIMO
over the UL.
It is possible to transmit/receive simultaneously 
to/from up to 74 stations over the DL/UL 
respectively. 

Finally, in this paper we assume an unreliable channel and therefore
the use of the IEEE 802.11 Automatic-ReQuest response (ARQ)
protocol. We do not specify this protocol here. The interested
reader can find a description of this protocol in e.g. paper~\cite{SA20}.

\section{Model}

\subsection{IEEE 802.11ax (HE) TCP aware scheduling strategies}

\subsubsection{IEEE 802.11ax (HE) DL unreliable channel, simultaneous Multi User unidirectional TCP operation mode}

\begin{figure}
\vskip 15cm
\includegraphics{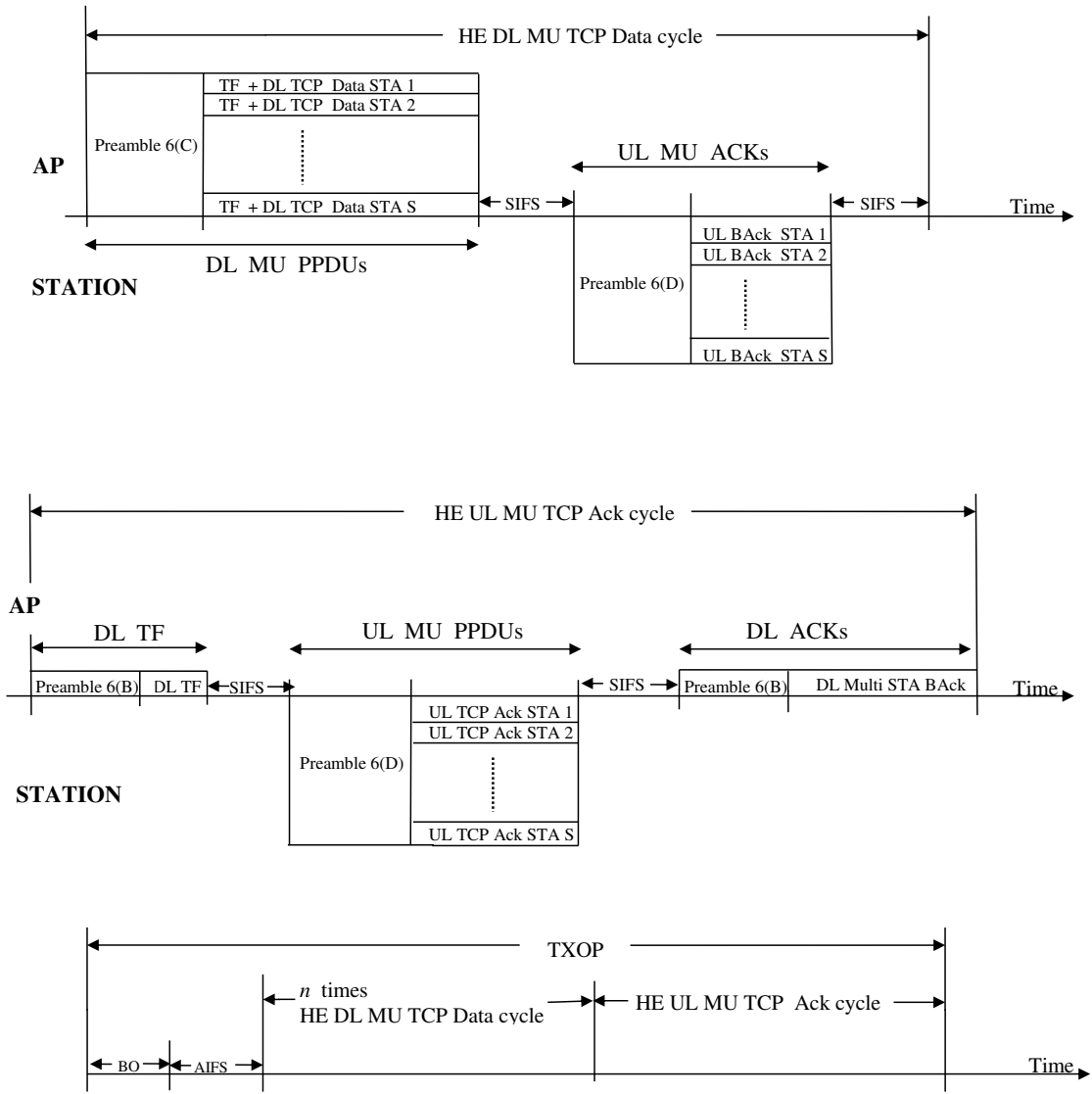}
\caption{The HE DL unreliable channel, simultaneous Multi User unidirectional TCP operation mode.}
\label{fig:mu}
\end{figure}

In the HE DL unreliable channel, simultaneous MU unidirectional
TCP, the AP transmits TCP
Data segments to several stations simultaneously
and receives TCP Ack segments
from the stations simultaneously.
We assume the following
11ax operation mode
where simultaneous DL TCP Data segments are
sent by the AP to multiple stations
in the same PPDU using MU mode, and the TCP Ack 
segments are sent simultaneously by the
stations in one PPDU
by again using
MU mode under unreliable channel conditions as
is illustrated in Figure~\ref{fig:mu}.

In the proposed operation mode,
after waiting
the BackOff and AIFS intervals, the AP
receives an air access and starts a TXOP by transmitting
a DL HE MU PPDU frame containing A-MPDU frames
which contain TCP Data segments,
to a given set of
stations simultaneously using MU mode. 
In the DL PPDU the AP transmits one A-MPDU
frame to every station in the MU stations' set, 
i.e. group of stations. Every A-MPDU contains
several MPDU frames and each MPDU contains several
MSDUs, which are TCP Data segments.
The MPDU frames in an A-MPDU are acknowledged by a BAck frame
from their receiver and all the BAck frames
are also transmitted simultaneously by the
stations using HE UL MU mode.
After the AP receives the BAck frames
the so-called
HE DL MU TCP Data cycle ends
and such a Data cycle
can now repeat itself several times. 

In order to reduce the delay one
needs to transmit to or receive from
a group of stations simultaneously, and
therefore the AP needs to
allocate Resource Units (RU)
per served station.
A RU is a subchannel and 
RU allocation is done for the DL for TCP Data segments
and for the UL for the BAck frames and the TCP Ack segments'
transmissions based on the traffic profile and channel
conditions.
According to the 11ax protocol,
the AP signals the stations when and how to transmit, i.e.
their UL RU allocation,
by a  Trigger Frame (TF) control frame.
For the BAck transmissions the AP transmits a
unicast Trigger
Frame (TF) to every station that contains the UL RU allocation. 
This TF MPDU
is aggregated with the DL Data MPDUs
that the AP transmits to a station in the DL HE MU A-MPDU
frame.

At the end of the last HE DL MU TCP Data cycle the AP
initiates a HE UL MU TCP Ack cycle by
transmitting the broadcast
Trigger Frame (TF) to the MU stations' set, i.e. group of
stations. 
This TF solicits MU UL Data transmissions
that contain TCP Ack transmissions from the
stations to the AP via the same unreliable channel.
In this UL transmission
the stations transmit TCP Ack segments
using UL HE MU A-MPDU frames using
HE triggered based MU PPDU per station. Every station transmits
its TCP Ack segments in a different UL HE MU A-MPDU frame.
The AP acknowledges the stations' UL HE MU A-MPDU
frames by generating and transmitting a single DL Multi Station BAck
(M-BA) frame.
The AP selects the UL TCP Ack transmission parameters
in order to increase the transmission efficiency and reliability.
The dominant parameters are the MCS, the number of spatial
streams to use and the number
of TCP Ack segments to include in UL triggered based MPDUs.
At this stage the HE UL MU TCP Ack cycle ends 
and a new series of HE DL MU TCP Data cycle(s) and
HE UL MU TCP Ack cycle begin.

Since the AP has full control over the channel
access activity 
we can assume that
there are no collisions
by e.g. increasing
the size of the congestion interval
from which the stations choose their HE MU Extended Distributed Coordination
Function (EDCF) BackOff 
parameter set, and/or the AP
can use the new 11ax MU air access restrictions' timers
that can stop a station from transmitting.

\subsubsection{TCP performance optimization}

In order to maximize the Goodput while minimizing the
round trip TCP delay, we propose TCP-aware
scheduling strategies that optimize the TXOP formation.
For this purpose one needs to
analyze the TXOP structure and the format and parameters
of the various frames
in the operation mode, i.e. the TXOP
formation optimization is done by
deciding how many MPDUs
containing TCP Acks are transmitted by
a station in its A-MPDU, how many TCP Acks
there are in an MPDU, 
what is the MCS and Number of Spatial Streams (NSS)
to be used by the stations,
how many A-MPDUs that contain TCP Data
segments are transmitted in a TXOP,
how many MPDUs there
are in such A-MPDUs and
how many TCP Data segments there are in an MPDU.
The selection is done based on the traffic profile
and the channels' conditions.

Based on the current 11ax product practical implementation
and in order to protect the
vulnerable portion of the TXOP, i.e. the UL TCP Ack transmission,
we first manage that the transmission error probability
of MPDUs containing TCP Acks is negligible compared to that
of MPDUs containing TCP Data segments. 
The negligible UL TCP Ack transmission
error probability is achieved
by using either a lower MCS and/or by shorter MPDUs (less TCP Acks).
Therefore, in this paper we set
the reception error of MPDUs containing UL TCP Acks
as practically $\sim 0$, and much smaller than
the reception error of MPDUs containing TCP Data segments,
i.e. $P_{error}(TCP\ Ack) << P_{error}(TCP\ Data)$,
by choosing for the UL the highest indexed MCS in which
BER$=$0.

In order to optimize the Goodput by
increasing 
the transmission efficiency further, there is
a need to minimize overhead 
as much as possible
when transmitting the TCP Acks.
Therefore it is most efficient
to transmit a given number
of TCP Ack segments in a minimal
number of A-MPDU frames,
given the maximum
number of such segments that
can be sent in one MPDU by the 11ax frame lengths and PPDU duration,
and the 0 (zero) target reception error probability in this paper.
Let an MPDU that contains
as many TCP Acks as possible 
according to the above criteria be denoted
a Full
MPDU. A Partial MPDU is one that
contains a smaller number of TCP Acks.
The A-MPDU containing
TCP Acks shall contain as many 
Full MPDUs
with possibly
one Partial MPDU, given the 11ax
protocols' limits which are the number
of MPDUs that cannot be
larger than 256, which is the 11ax
Block Ack (BA) window size and the transmission
duration time of the MAC PPDU containing the A-MPDU that
can be at
most $5.484 ms$.
This optimization derives the number of TCP Data
segments to be transmitted in a TXOP, as we explain in Section
3.2.2 .

Concerning the number of TCP Data segments
to be transmitted in one MPDU, we use the
{\it local throughput optimization strategy} 
in unreliable channels
which maximizes the DL MPDUs throughput in the
following way.
Say X TCP Data segments are transmitted in an DL MPDU containing
TCP Data segments.  Let $Bits(X)$ and $Time(X)$ be the length
in bits
of the DL MPDU containing these X TCP Data segments, and its
transmission time respectively. Let $L_{Data}$ be
the length in bits of a TCP Data segment and BER be the
Bit Error Rate. Then, $X \cdot L_{Data} \cdot (1-BER)^{Bits(X)}$
is the average number of TCP Data bits being transmitted
successfully in the MPDU.
The local throughput of the MPDU is given by:

\begin{equation}
U(X)=
\frac
{X \cdot L_{Data} \cdot (1-BER)^{Bits(X)}}
{Time(X)}
\label{equ:1}
\end{equation}

One then finds the optimal integer X that maximizes the local throughput, 
and say it is X$^{*}$, i.e.:

\begin{equation}
X^{*} = \max_{X} U(X)
\end{equation}

Then, in every MPDU one transmits X$^{*}$ TCP 
Data segments.
If a smaller number than X$^{*}$ TCP Data
segments are (left) to be transmitted
in an A-MPDU, say $Y < X^{*}$ Data segments,
these Y TCP Data segments are transmitted in one DL MPDU.

\subsubsection{Proposed TCP-aware scheduling strategies}

In relation to every TCP connection
the target is to reach an optimized working point where the number
of TCP Data segments transmitted and acknowledged
at the MAC layer, but-not-yet-acknowledged 
at the TCP layer, to
be at least the optimized number of  TCP Acks to
be transmitted at the end of the TXOP.
Let $S$ be the maximum number of TCP Acks that
a station can transmit 
in one A-MPDU. Let $Base$ be the serial
number of the first TCP Data segment
that was transmitted by the AP but not-yet-acknowledged at
the TCP layer.
If the AP was transmitting to one station only, 
then it was continuing to transmit
DL A-MPDUs until the number
of TCP Data segments 'on the fly', i.e. that
were already transmitted and acknowledged
at the MAC layer, but-not-yet-acknowledged at the TCP layer, is
at least $S$ in a row starting from $Base$.
After this number is reached, the AP enables the receiving
station to transmit its UL A-MPDU containing the
TCP Acks. Notice that the AP knows which
TCP Data segments arrived successfully or in error at the
receiver at the MAC layer by the BAck frames that it receives
from the station after each DL A-MPDU transmission.

However, since the AP transmits to a set of stations simultaneously,
it can be the case that not all the
receiving stations reach the point where they have at least
$S$ TCP Acks per station to transmit simultaneously. Therefore, we check
three possible scheduling strategies which
differ in their termination criteria:

\begin{enumerate}

\item
TCP-aware
scheduling strategy 1 - minimal response time:
The AP transmits until at least one station has $S$ TCP Acks to transmit.

\item
TCP-aware
scheduling strategy 2 - target response time:
The AP transmits until all the stations have $S$ TCP Acks per
station to transmit.

\item
TCP-aware
scheduling strategy 3 - max Goodput based response time:
Given that the AP transmits to $N$ stations then it transmits
until all the stations together have at least $Load \cdot N \cdot S$
TCP Acks to transmit, where $0 < Load \le 1$.

\end{enumerate}

Together with the definition of the above scheduling
strategies, we also assume that the AP transmits
as much TCP Data segments as it can in an A-MPDU. 
When the AP transmits an A-MPDU according to the
above strategies, it is sometimes left with the capability
to transmit in the A-MPDU more TCP Data segments
than the strategy requires, and the AP does so.

\subsection{Parameters' values}

\subsubsection{RU allocation}

We assume the 5GHz band and a 160MHz AP operation bandwidth
that is divided into either
2 RUs of 80MHz each,
4 RUs of 40MHz each, or into
8 RUs of 20MHz each. In the case of a 160MHz RU allocation
the AP communicates
with 4 stations using
one spatial stream per station.
In the case of 2 RUs of 80MHz each, in every
80MHz RU the AP communicates with 4 stations, using
one spatial stream per station. Thus, in the
two RUs the AP communicates with 8 stations.
The same holds for the 40MHz and 20MHz RUs allocations,
in which the AP communicates with 16 and 32 stations
respectively. 
When 4 stations are served in the system
the 160MHz operation bandwidth is used in MU-MIMO.
For  a larger number of served 
stations MU-MIMO+OFDMA is used. 
The stations transmit to the AP over
the UL in a symmetrical way to that of the
AP over the DL. 

The TF and the Multi Station BAck frames are
transmitted using the legacy mode and the
PHY rate 
$R_{legacy}$ is set to the largest basic rate that
is smaller or equal to
the TCP Data/Ack segments' transmission rate $R_{TCP}$.

In Table~\ref{tab:phyrates} we show the PHY rates and the preambles
used in the various MCSs and
in all cases of the number of stations $S$, i.e. 
$S= 4, 8, 16$ and 32.

\begin{table}
\caption{\label{tab:phyrates}{The PHY rates and preambles in IEEE 802.11ax . 160, 80, 40 and 20 MHz channels are assumed, with 4 spatial streams in each for 4, 8, 16 and 32 stations respectively. The TF and BAck
transmissions are conducted at the basic rate set. }}
\vspace{3 mm}
\tiny
\center
\begin{tabular}{|r|c|c|c|c|c|c|c|}  \hline
     & \multicolumn{2}{|c|}1  & \multicolumn{2}{|c|}2 & &
       \multicolumn{2}{|c|}3  \\ \cline{2-8}
     & \multicolumn{2}{c|}{MU UL data}& \multicolumn{2}{c|}{MU DL data}&  & \multicolumn{2}{c|}{DL TF/Multi Station BAck}  \\ 
& \multicolumn{2}{c|}{transmission rate} & \multicolumn{2}{c|}{transmission rate} &  & \multicolumn{2}{c|}{transmission rate}  \\ \hline
     &  PHY Rate & Preamble & PHY Rate & Preamble & 
        & PHY Rate (legacy) & Preamble \\
MCS  &  (Mbps per 1 SS)   & ($\mu s$)  & (Mbps per 1 SS) & ($\mu s$)
     & & (Mbps)   & ($\mu s$) \\
     &  GI$=1.6 \mu s$   & & GI$=0.8 \mu s$ &
      &  & GI$=0.8 \mu s$ &  \\ \hline
      \multicolumn{8}{|c|}{4 stations}  \\ \hline
 0   &  68.1   & 64.8 & 72.1 & 72.8 && 48.0 & 20.0 \\ 
 1   &  136.1  & 64.8 &144.1 & 72.8 && 48.0 & 20.0 \\ 
 2   &  204.2  & 64.8 &216.2 & 68.8 && 48.0 & 20.0 \\ 
 3   &  272.2  & 64.8 &288.2 & 68.8 && 48.0 & 20.0 \\ 
 4   &  408.3  & 64.8 &432.4 & 68.8 && 48.0 & 20.0 \\ 
 5   &  544.4  & 64.8 &576.5 & 68.8 && 48.0 & 20.0 \\ 
 6   &  612.5  & 64.8 &648.5 & 68.8 && 48.0 & 20.0 \\ 
 7   &  680.6  & 64.8 &720.6 & 68.8 && 48.0 & 20.0 \\ 
8    &  816.7  & 64.8 &864.7 & 68.8 && 48.0 & 20.0 \\ 
 9   &  907.4  & 64.8 &960.7 & 68.8 && 48.0 & 20.0 \\ 
10   & 1020.8  & 64.8 &1080.4 & 68.8 && 48.0 & 20.0 \\ 
11   & 1134.2  & 64.8 &1201.0 & 68.8 && 48.0 & 20.0 \\ \hline 
      \multicolumn{8}{|c|}{8 stations}  \\ \hline
 0   &  34.0 & 64.8 &36.0 & 76.8 && 36.0 & 20.0  \\ 
 1   &  68.1 & 64.8 &72.1 & 76.8 && 48.0 & 20.0  \\ 
 2   & 102.1 & 64.8 &108.1 & 72.8 && 48.0 & 20.0  \\ 
 3   & 136.1 & 64.8 &144.1 & 72.8 && 48.0 & 20.0  \\ 
 4   & 204.2 & 64.8 &216.2 & 68.8 && 48.0 & 20.0  \\ 
 5   & 272.2 & 64.8 &288.2 & 68.8 && 48.0 & 20.0  \\ 
 6   & 306.3 & 64.8 &324.3 & 68.8 && 48.0 & 20.0  \\ 
 7   & 340.3 & 64.8 &360.3 & 68.8 && 48.0 & 20.0  \\ 
8    & 408.3 & 64.8 &432.4 & 68.8 && 48.0 & 20.0  \\ 
 9   & 453.7 & 64.8 &480.4 & 68.8 && 48.0 & 20.0  \\ 
10   & 510.4 & 64.8 &540.4 & 68.8 && 48.0 & 20.0  \\
11   & 567.1 & 64.8 &600.4 & 68.8 && 48.0 & 20.0  \\ \hline 
     \multicolumn{8}{|c|}{16 stations}  \\ \hline
 0   &  16.3 & 64.8 &17.2 & 84.8 && 12.0 & 20.0 \\ 
 1   &  32.5 & 64.8 &34.4 & 84.8 && 12.0 & 20.0 \\ 
 2   &  48.8 & 64.8 &51.6 & 76.8 && 24.0 & 20.0 \\ 
 3   &  65.0 & 64.8 &68.8 & 76.8 && 48.0 & 20.0 \\ 
 4   &  97.5 & 64.8 &103.2 & 72.8 && 48.0 & 20.0 \\ 
 5   & 130.0 & 64.8 &137.6 & 72.8 && 48.0 & 20.0 \\ 
 6   & 146.3 & 64.8 &154.9 & 72.8 && 48.0 & 20.0 \\ 
 7   & 162.5 & 64.8 &172.1 & 72.8 && 48.0 & 20.0 \\ 
8    & 195.0 & 64.8 &206.5 & 72.8 && 48.0 & 20.0 \\ 
 9   & 216.7 & 64.8 &229.4 & 72.8 && 48.0 & 20.0 \\ 
10   & 243.8 & 64.8 &258.1 & 72.8 && 48.0 & 20.0 \\ 
11   & 270.8 & 64.8 &286.8 & 72.8 && 48.0 & 20.0 \\ \hline 
\end{tabular} 
\end{table}

\addtocounter{table}{-1}

\begin{table}
\caption{\label{tab:phyrates1}{(cont.)}}
\vspace{3 mm}
\tiny
\center
\begin{tabular}{|r|c|c|c|c|c|c|c|}  \hline
     & \multicolumn{2}{|c|}1  & \multicolumn{2}{|c|}2 & &
       \multicolumn{2}{|c|}3  \\ \cline{2-8}
     & \multicolumn{2}{c|}{MU UL data}& \multicolumn{2}{c|}{MU DL data}&  & \multicolumn{2}{c|}{DL TF/Multi Station BAck}  \\ 
& \multicolumn{2}{c|}{transmission rate} & \multicolumn{2}{c|}{transmission rate} &  & \multicolumn{2}{c|}{transmission rate}  \\ \hline
     &  PHY Rate & Preamble & PHY Rate & Preamble & 
        & PHY Rate (legacy) & Preamble \\
MCS  &  (Mbps per 1 SS)   & ($\mu s$)  & (Mbps per 1 SS) & ($\mu s$)
     & & (Mbps)   & ($\mu s$) \\
     & GI$=1.6 \mu s$ & & GI$=0.8 \mu s$ &
      &  & GI$=0.8 \mu s$ &  \\ \hline
     \multicolumn{8}{|c|}{32 stations} \\ \hline
 0   &   8.1 & 64.8 &8.6& 104.8 &&  6.0 & 20.0 \\ 
 1   &  16.3 & 64.8 &17.2& 104.8 && 12.0 & 20.0 \\ 
 2   &  24.4 & 64.8 &25.8& 84.8 && 24.0 & 20.0 \\ 
 3   &  32.5 & 64.8 &34.4& 84.8 && 24.0 & 20.0 \\ 
 4   &  48.8 & 64.8 &51.6& 80.8 && 48.0 & 20.0 \\ 
 5   &  65.0 & 64.8 &68.8& 80.8 && 48.0 & 20.0 \\ 
 6   &  73.1 & 64.8 &77.4& 80.8 && 48.0 & 20.0 \\ 
 7   &  81.3 & 64.8 &86.0& 80.8 && 48.0 & 20.0 \\ 
8    &  97.5 & 64.8 &103.2& 80.8 && 48.0 & 20.0 \\ 
 9   & 108.3 & 64.8 &114.7& 80.8 && 48.0 & 20.0 \\ 
10   & 121.9 & 64.8 &129.0& 80.8 && 48.0 & 20.0 \\ 
11   & 135.4 & 64.8 &143.4& 80.8 && 48.0 & 20.0 \\ \hline 
%      \multicolumn{8}{|c|}{64 stations}  \\ \hline
%  0   &  3.5 & 64.8 &3.8& 136.8 &&  6.0 & 20.0 \\ 
%  1   &  7.1 & 64.8 &7.5& 136.8 &&  6.0 & 20.0 \\ 
%  2   & 10.6 & 64.8 &11.3& 100.8 &&  9.0 & 20.0 \\ 
%  3   & 14.2 & 64.8 &15.0& 100.8 && 12.0 & 20.0 \\ 
%  4   & 21.3 & 64.8 &22.5& 88.8 && 18.0 & 20.0 \\ 
%  5   & 28.3 & 64.8 &30.0& 88.8 && 24.0 & 20.0 \\ 
%  6   & 31.9 & 64.8 &33.8& 88.8 && 24.0 & 20.0 \\ 
%  7   & 35.4 & 64.8 &37.5& 88.8 && 24.0 & 20.0 \\ 
%  8   & 42.5 & 64.8 &45.0& 88.8 && 36.0 & 20.0 \\ 
%  9   & 47.2 & 64.8 &50.0& 88.8 && 36.0 & 20.0 \\ 
% 10   &  N/A &  N/A &N/A& N/A &&  N/A &  N/A \\ 
% 11   &  N/A &  N/A &N/A& N/A &&  N/A &  N/A \\ \hline 
\end{tabular}  
\end{table}

In this paper we assume unreliable channels
and need to consider the relation between the
Bit Error Rate (BER) to the bandwidth of the
channel in use, the MCS in use and the Signal-to-Noise-Ratio (SNR).
Using the IEEE official channel mode description~\cite{IEEEber}
one can find for every RU's bandwidth
and MCS the relation between the SNR and BER, assuming
that the AP is communicating with every station over
one spatial stream. We demonstrate the above
in the Appendix.

\subsubsection{Channel Access and frames' size}

We assume the IEEE 802.11
Best Effort Access Category
in which $AIFS=43 \mu s$ for
the AP and $52 \mu s$ for a station, $SIFS=16 \mu s$ and $CW_{min}=16$
for the transmissions of the AP.
Recall that
we assume there are
no collisions between the AP and the stations
because we use the 11ax triggered based AP operation.
The BackOff interval is a random number
chosen uniformly from
the range $[0,....,CW_{min}-1]$. 
Since we consider a very 'large' number
of transmissions from the AP
we take the BackOff average
value of  
$\ceil{\frac{CW_{min}-1}{2}}$, and the average BackOff interval 
for the AP is
$\ceil{\frac{CW_{min}-1}{2}} \cdot SlotTime$
which equals $67.5 \mu s$ for a $SlotTime= 9 \mu s$.

Concerning the transmission in 11ax mode,
an OFDM symbol
is $12.8 \mu s$. In the DL
we assume a GI of $0.8 \mu s$ and therefore the symbol
in this direction is $13.6 \mu s$. In the UL MU we assume
a GI of $1.6 \mu s$ and therefore the symbol in this
direction is $14.4 \mu s$. The UL GI is $1.6 \mu s$ 
due to UL arrival time variants.
When considering transmissions in legacy mode, the symbol
is $4 \mu s$ containing a GI of $0.8 \mu s$.

We assume that the MAC Header field
is of 28 bytes and
the Frame Control Sequence (FCS) field is of 4 bytes.
Finally, we assume
TCP Data segments of $L_{DATA}=1460$ and
$208$ bytes. Therefore, the resulting
MSDUs' lengths are $L^{'}_{DATA}=1508$ and $256$
bytes respectively ( 20 bytes of TCP header
plus 20 bytes of IP header plus
8 bytes of LLC SNAP are added ).
Together with the SubHeader field and rounding
to an integral multiple of 4 bytes, every MSDU is now
of $Len^{D}=1524$ and $272$ bytes respectively. 
Due to the limit
of 11454 bytes on the MPDU size, 7
and 42 such MSDUs are possible respectively in one MPDU.

The TCP receiver transmits TCP Acks. Every MSDU containing
a TCP Ack
is of $L^{'}_{Ack}=48$ bytes 
( 20 bytes of TCP Header + 20 bytes of IP header +
8 bytes of LLC SNAP ). Adding 14 bytes of the SubHeader
field and rounding to an integral multiple of 4 bytes, every
MSDU is of $Len^{A}=64$ bytes, and every
single MPDU, again due to the size limit
of 11454 bytes, can contain up to 178 MSDUs (TCP Acks).
Thus, the receiver can transmit
up to $N_{MAX}=256 \cdot 178$ TCP
Acks in a single HE UL A-MPDU frame.

\section{Performance results}

In the following we show simulation performance results
for the 11ax TCP scheduling strategies we defined previously. 
All the simulation results addressed in this paper were
obtained by software that we wrote, and were validated
by NS-3 and 11ax products.
We checked
the performance as a function of the number of served stations
to which the AP transmits simultaneously, i.e. 4, 8, 16 and
32 stations.
In the following graphs' titles we use the number of
stations rather than the RU allocation,
to show the impact of the number of stations
on the results.

We show TCP Goodput
and Delay results. The TCP Goodput results show the
long term average number of TCP Data bits that are
successfully transmitted per second.
The related Delay results show the average TXOP length
as shown in the bottom
diagram of Figure~\ref{fig:mu}.
% The TXOP contains
% the BackOff interval, the AIFS interval, the length of the HE DL MU
% TCP Data cycles in the TXOP, the length of the HE
% UL MU TCP Ack cycle and the transmission of the various control
% frames.
This Delay definition is a measure
of the Round Trip Time (RTT) between the AP and the stations.

\begin{figure}
\vskip 16cm
\includegraphics{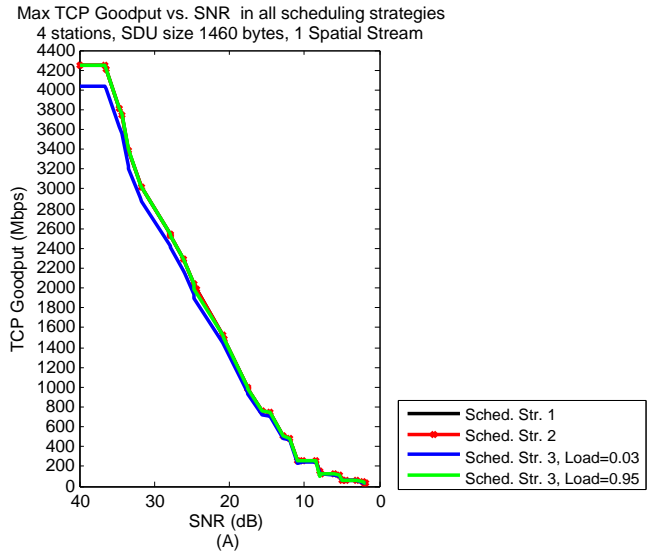}
\includegraphics{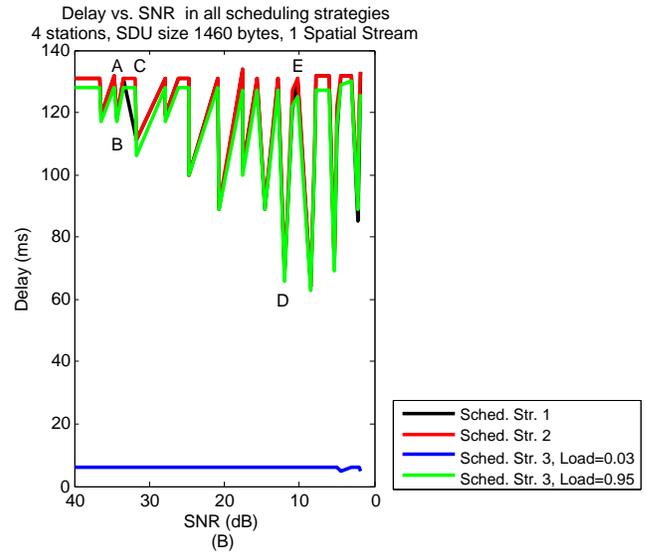}
\includegraphics{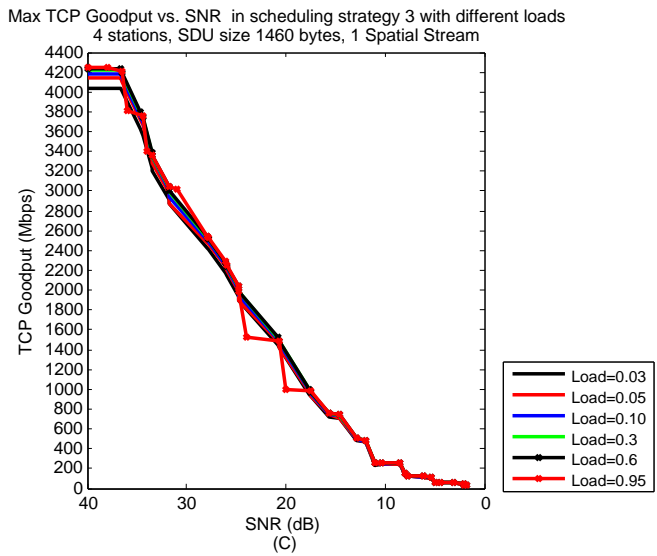}
\includegraphics{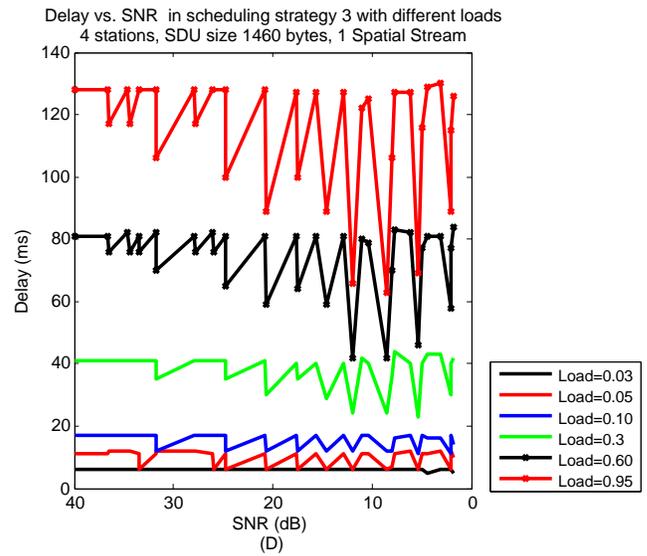}
\caption{TCP Goodput and Delay results for 4 stations and TCP Data segments of 1460 bytes, for the three scheduling strategies.}
\label{fig:160thrdelsnr}
\end{figure}

In Figure~\ref{fig:160thrdelsnr} we show the TCP Goodput
and Delay results for 4 stations, i.e. one 160MHz channel,
and for TCP Data segments of 1460 bytes. 
In Figures~\ref{fig:160thrdelsnr}(A)
and~\ref{fig:160thrdelsnr}(B) we show the maximum TCP Goodput and Delay results
respectively as a function of the SNR for scheduling
strategies 1, 2 and 3, where for scheduling strategy 3
we assume loads of 0.03 and 0.95 .

It is clear from the figures that scheduling strategy 3
with load=0.03 is the best to use in terms
of Delay optimization because it achieves about $92 \%$
of the maximum
TCP Goodput of the other strategies, but
with a delay of $6ms$, compared to about $120ms$ in the
other strategies which all have a similar performance.

Notice that the Delay curves in Figure~\ref{fig:160thrdelsnr}(B)
are swinging, except for the case of scheduling strategy 3 with
load=0.03 . The reasons for this swinging in the curves are subtle.
Notice points A and B in the curves. For point A, which corresponds
to SNR$=$36.6, one uses MCS11 over both
the DL and the UL with PHY rates of 1201.0 and 1134.2 Mbps
respectively, and both channels are reliable,
i.e. BER$=$0. On the other hand, point B corresponds to
SNR$=$36.5 where the BER over the DL is 0.0006 using MCS11.
Since we require a reliable UL channel one needs to use
MCS10 over the UL with a PHY rate of 1020.8 Mbps.
Thus, a smaller number of TCP Ack segments can be transmitted
in one HE UL MU TCP Ack cycle, and this results in a smaller
number of HE DL MU TCP Data cycles in the TXOP, and in a smaller
Delay.

In point C the Delay increases again. Point C corresponds to
SNR$=$34.7 where MCS10 is used over both the DL and the UL
with BER$=$0. The number of TCP Acks to transmit in a TXOP
remains as in point B but the DL PHY rate is smaller, which
causes the TXOP to be longer compared to point B.
Worth mentioning is also point D where the UL PHY rate becomes
very small, 68.1 Mbps when using MCS0. This small PHY rate causes a small
number of TCP Ack segments to be transmitted in a TXOP, and so
also to a small number of TCP Data segments in a TXOP. At this point
MCS1 is used over the DL with a PHY rate of 144.1 Mbps.
There is an increase in the Delay in point E because at this
point MCS0 is used over both the DL and the UL with a
PHY rate of 68.1 Mbps over both channels. This
results in a longer TXOP that is used
to transmit the same number of TCP Data/TCP Ack segments
as in point D.

There are no visible fluctuations in the curve for
scheduling strategy 3 with load$=$0.03 because the
number of TCP Data segments to transmit in a TXOP
is relatively small, and the differences in the PHY
rates used over the SNR range cause differences only 
in the order of a few tenth of $\mu s$, which are not
visible in the graph.

In Figures~\ref{fig:160thrdelsnr}(C) and~\ref{fig:160thrdelsnr} (D) we show
the sensitivity analysis of the load parameter in scheduling strategy
3 on the TCP Goodput and Delay results respectively.
One can see that up to load$=$0.1 the delays remain small
while for larger loads the delay increases significantly
without any gain in TCP Goodput, as can be observed from
Figure~\ref{fig:160thrdelsnr}(C).

\begin{figure}
\vskip 16cm
\includegraphics{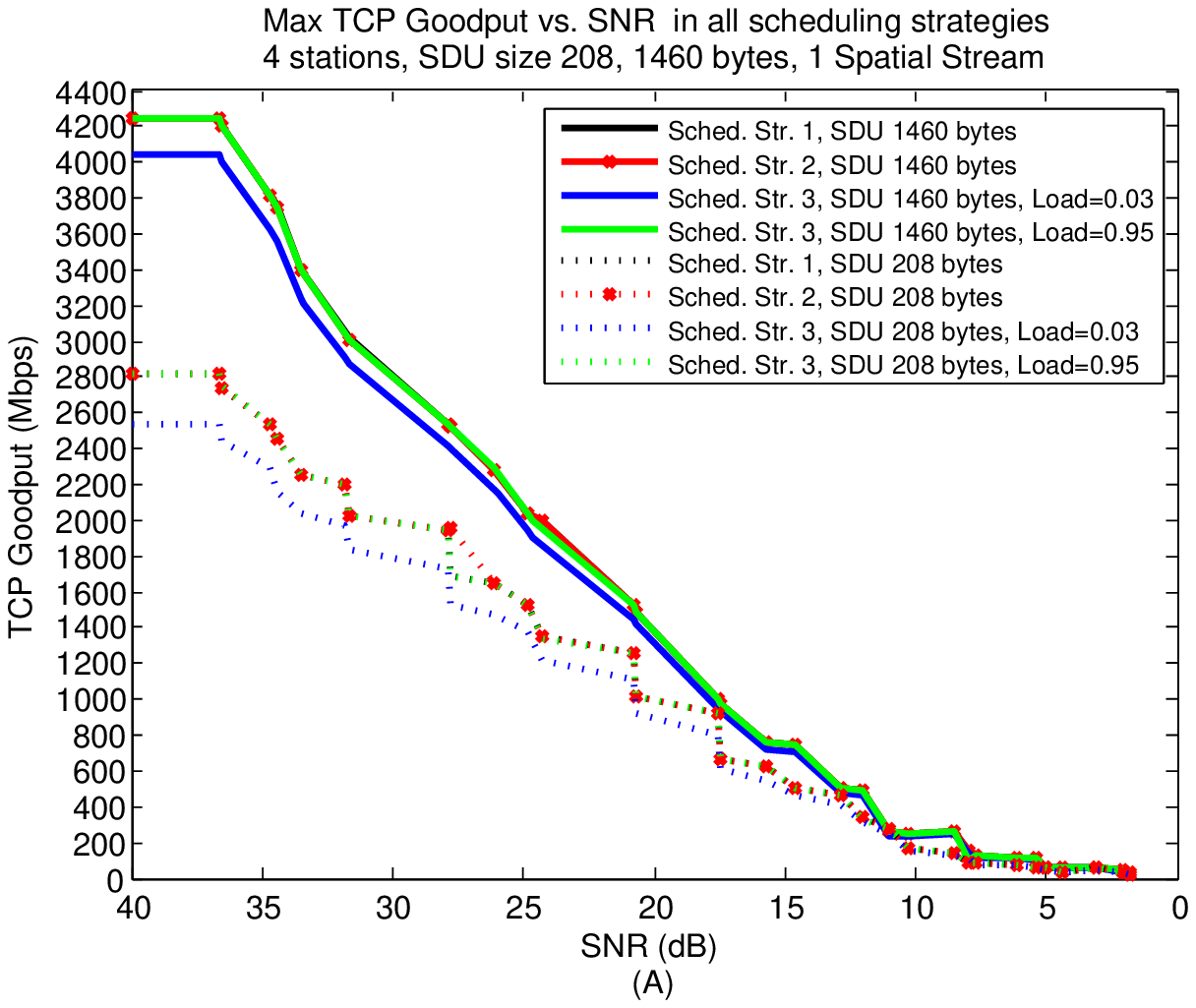}
\includegraphics{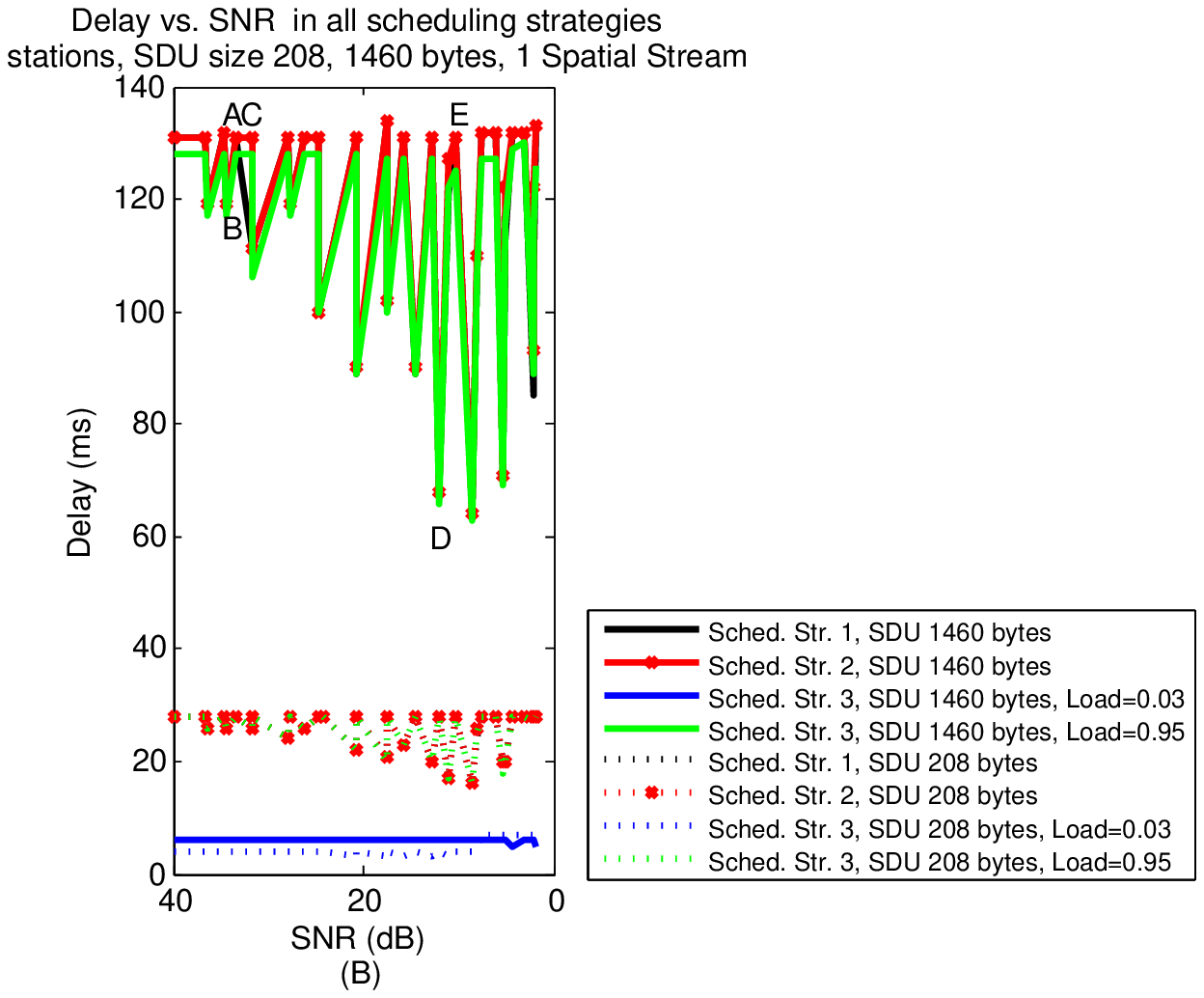}
\includegraphics{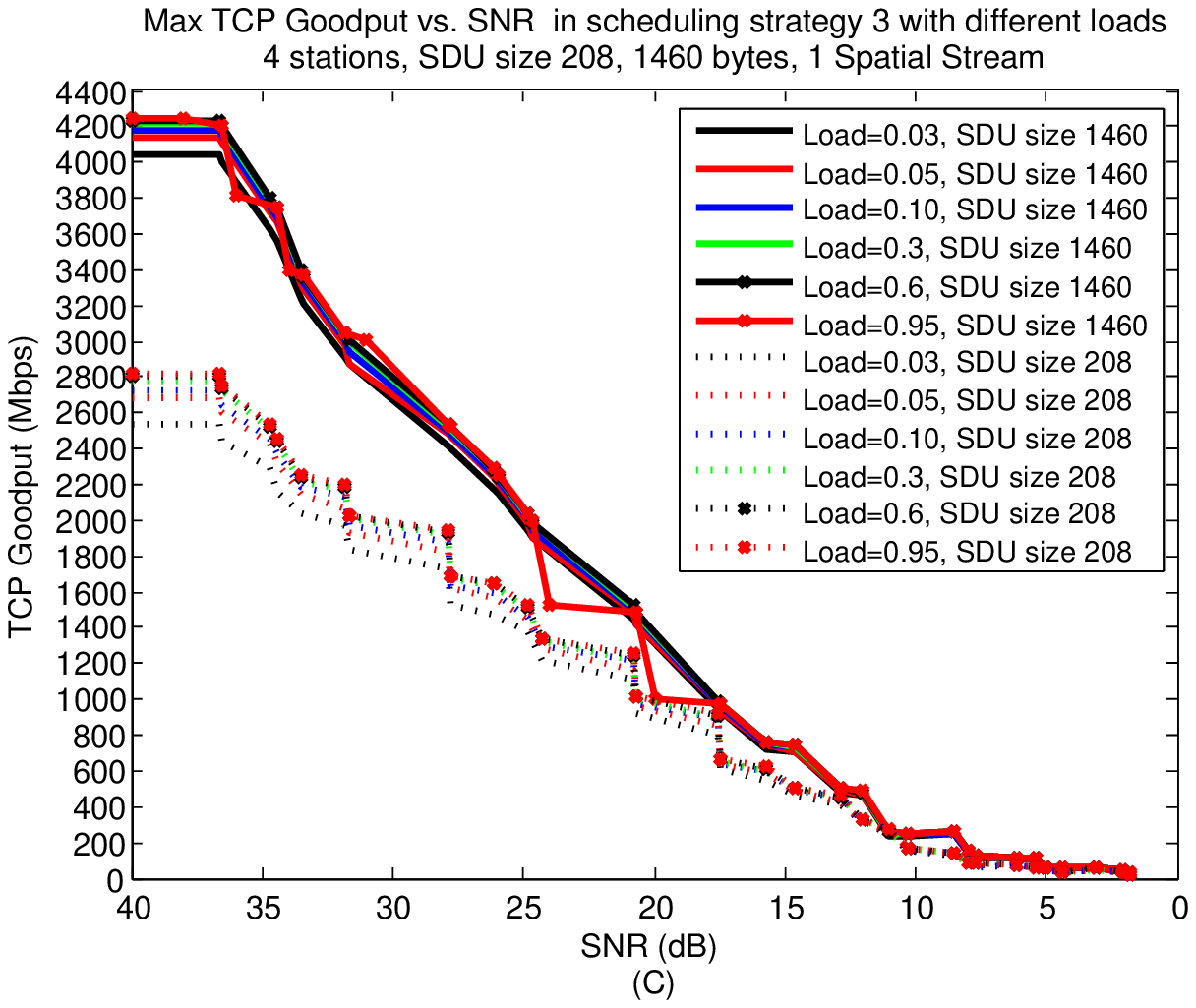}
\includegraphics{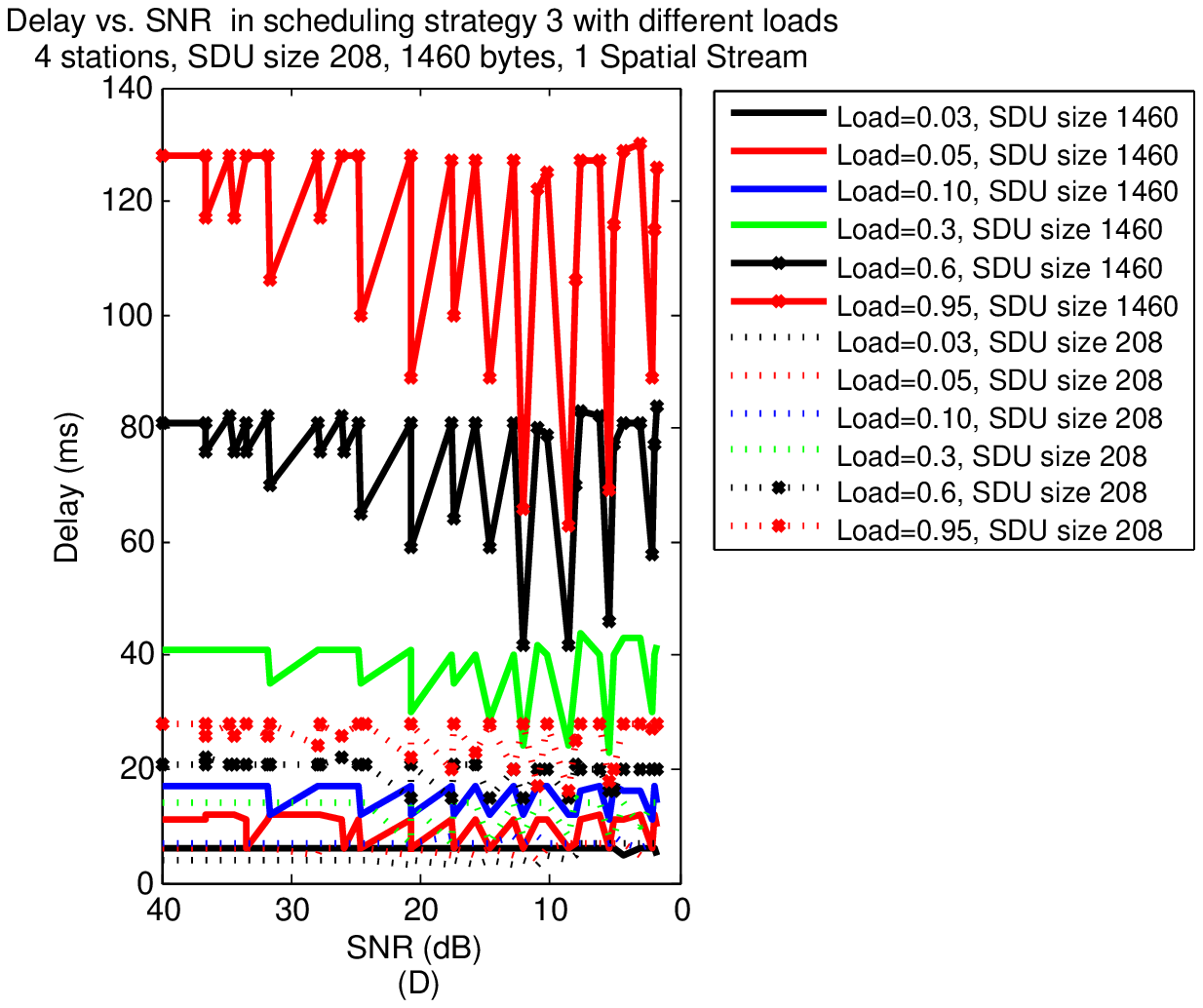}
\caption{TCP Goodput and Delay results for 4 stations and TCP Data segments of 208 and 1460 bytes, for the three scheduling strategies.}
\label{fig:160thrdelsnrc}
\end{figure}

In Figure~\ref{fig:160thrdelsnrc} we show the same results
as in Figure~\ref{fig:160thrdelsnr} but this time for TCP
Data segments of 208 and 1460 bytes together. For large SNR values
the TCP Goodput results for TCP Data segments
of 208 bytes
are smaller than those for 1460 bytes. In both cases approximately
the same number of TCP Ack segments are transmitted
in a TXOP, but in the case of the short TCP Data
segments the overhead (BackOff, AIFS etc.) is amortized
over a smaller number of TCP Data bits, resulting in a
lower TCP Goodput. 

As the SNR decreases 
the TCP Goodputs in both
segments' sizes converge. As the SNR decreases so
do the DL and UL PHY rates. This results in
long transmission times of the TCP Data segments
for both TCP Data segments' sizes and the
difference in length between 208 and 1460 bytes
becomes less dominant.

\begin{figure}
\vskip 16cm
\includegraphics{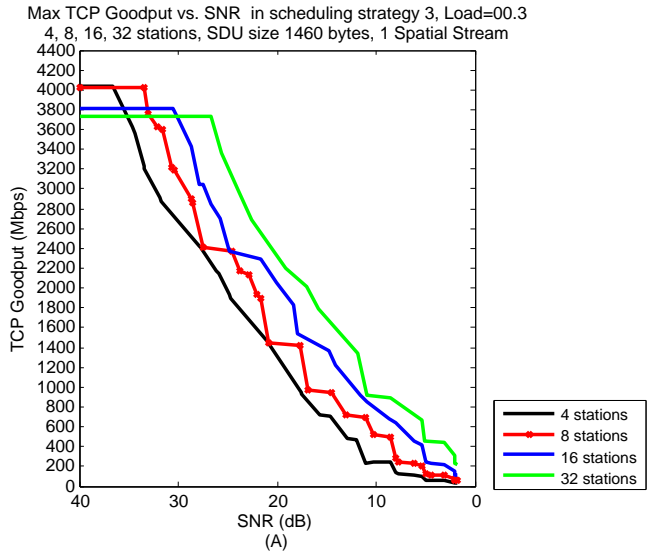}
\includegraphics{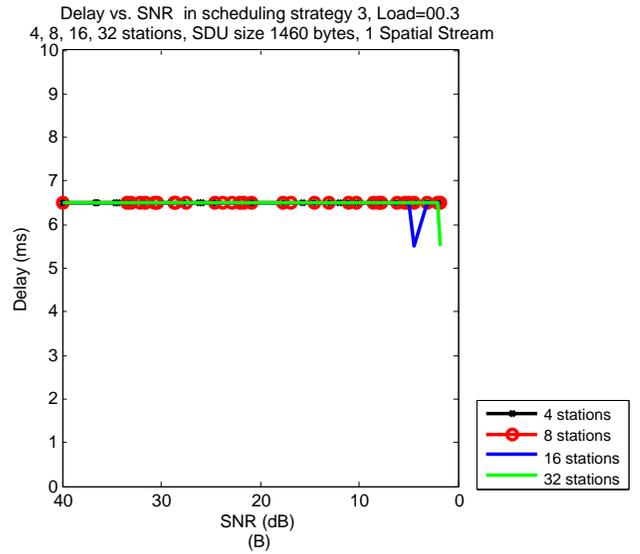}
\includegraphics{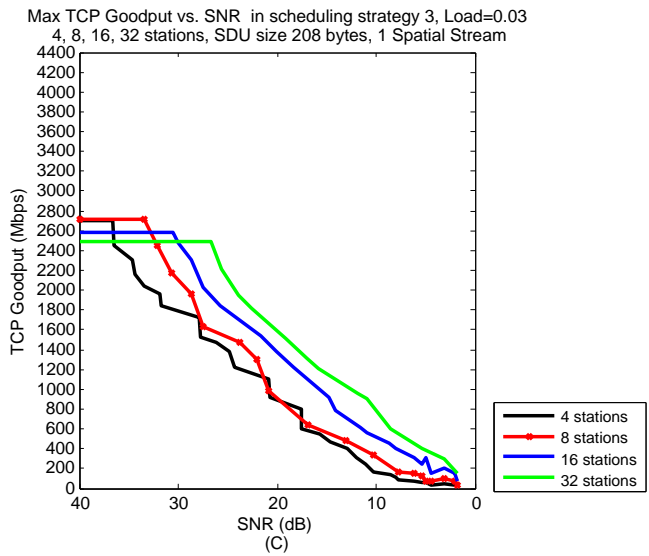}
\includegraphics{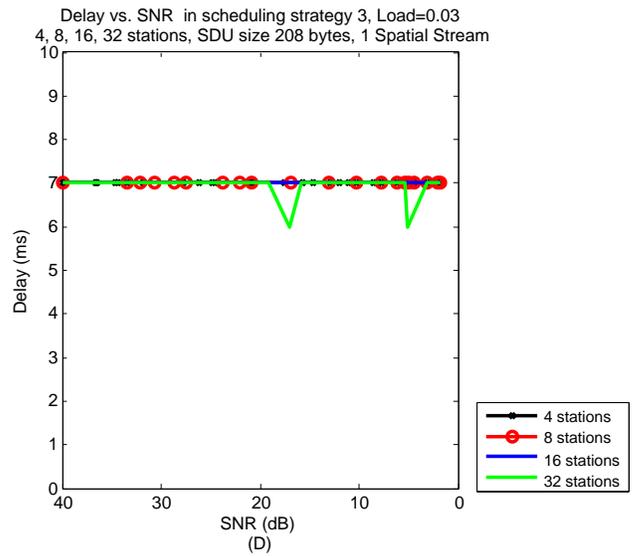}
\caption{TCP Goodput and Delay results for scheduling strategy 3 and
load $=$ 0.03, for 4, 8, 16 and 32 stations and TCP Data segments
of 208 and 1460 bytes.}
\label{fig:load003snr}
\end{figure}

Concerning the Delay metric, in general the TXOPs containing
TCP Data segments of 208 bytes are shorter compared
to those containing similar numbers of
TCP Data segments of 1460 bytes.
Therefore the delays are smaller in the former case.
An exception is scheduling strategy 3 with load$=$0.03, as we explain
in relation to
Figure~\ref{fig:load003snr} later.
Scheduling strategy 3 with load$=$0.03 is the most
efficient in both segments' sizes.

\begin{figure}
\vskip 16cm
\includegraphics{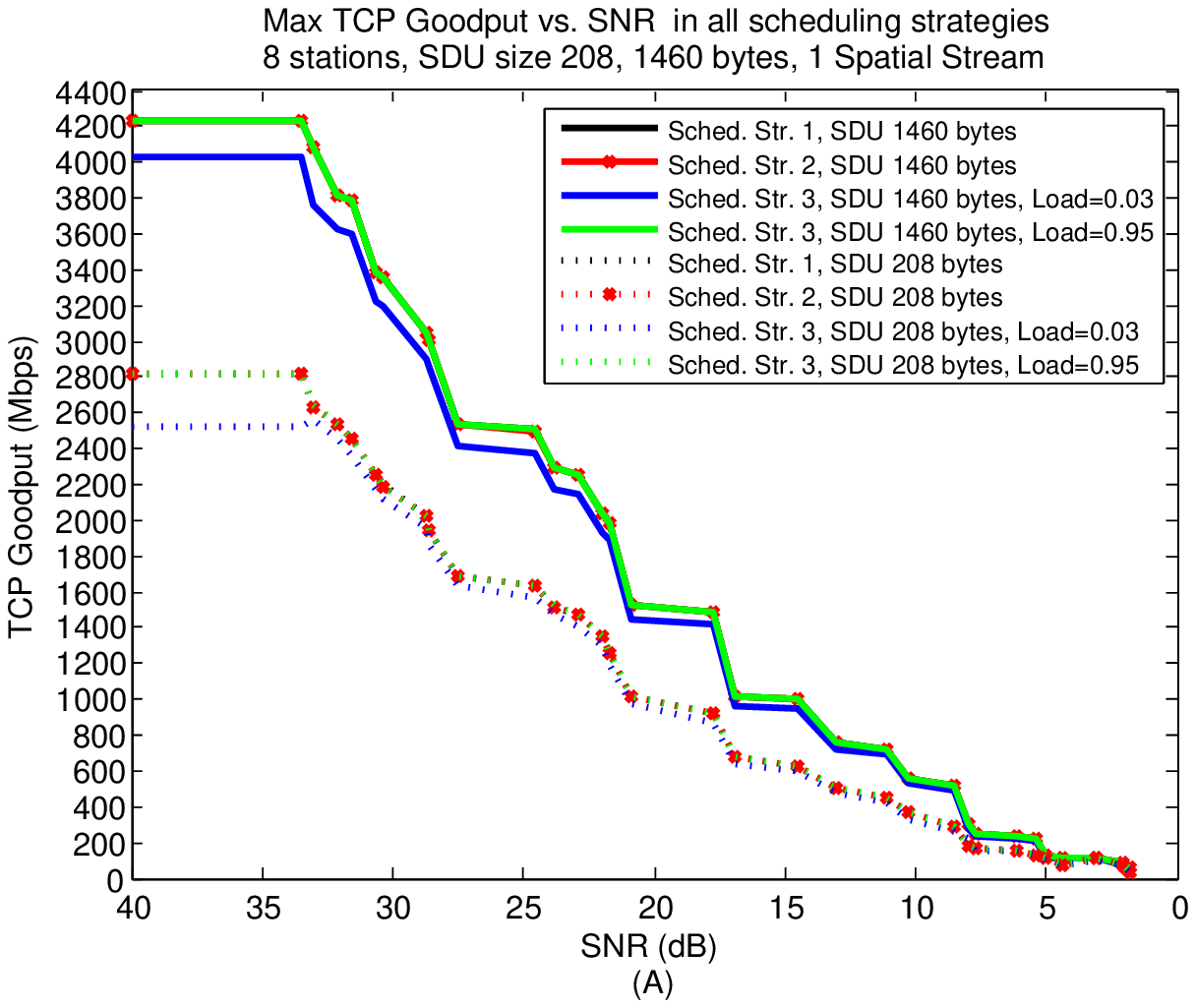}
\includegraphics{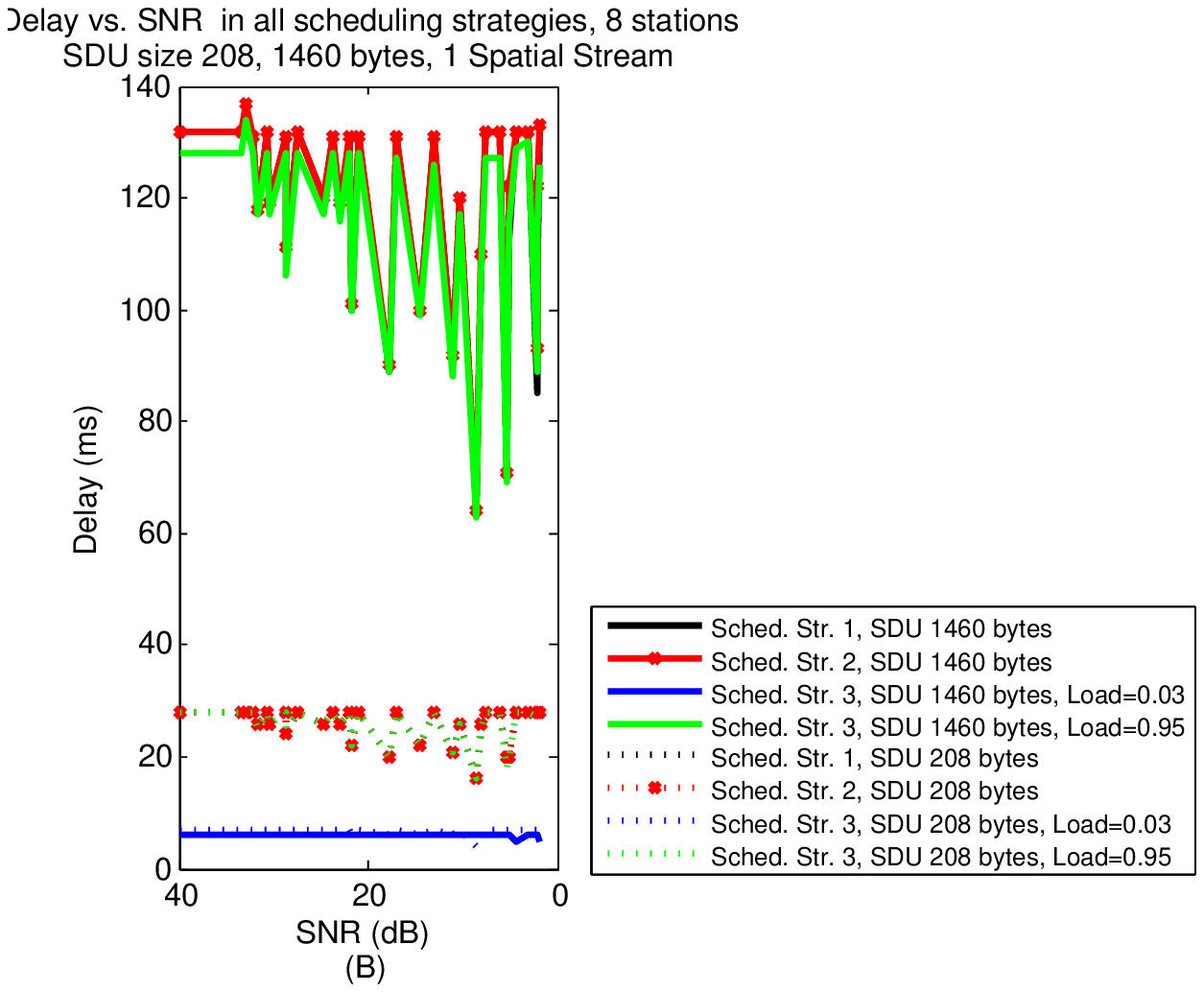}
\includegraphics{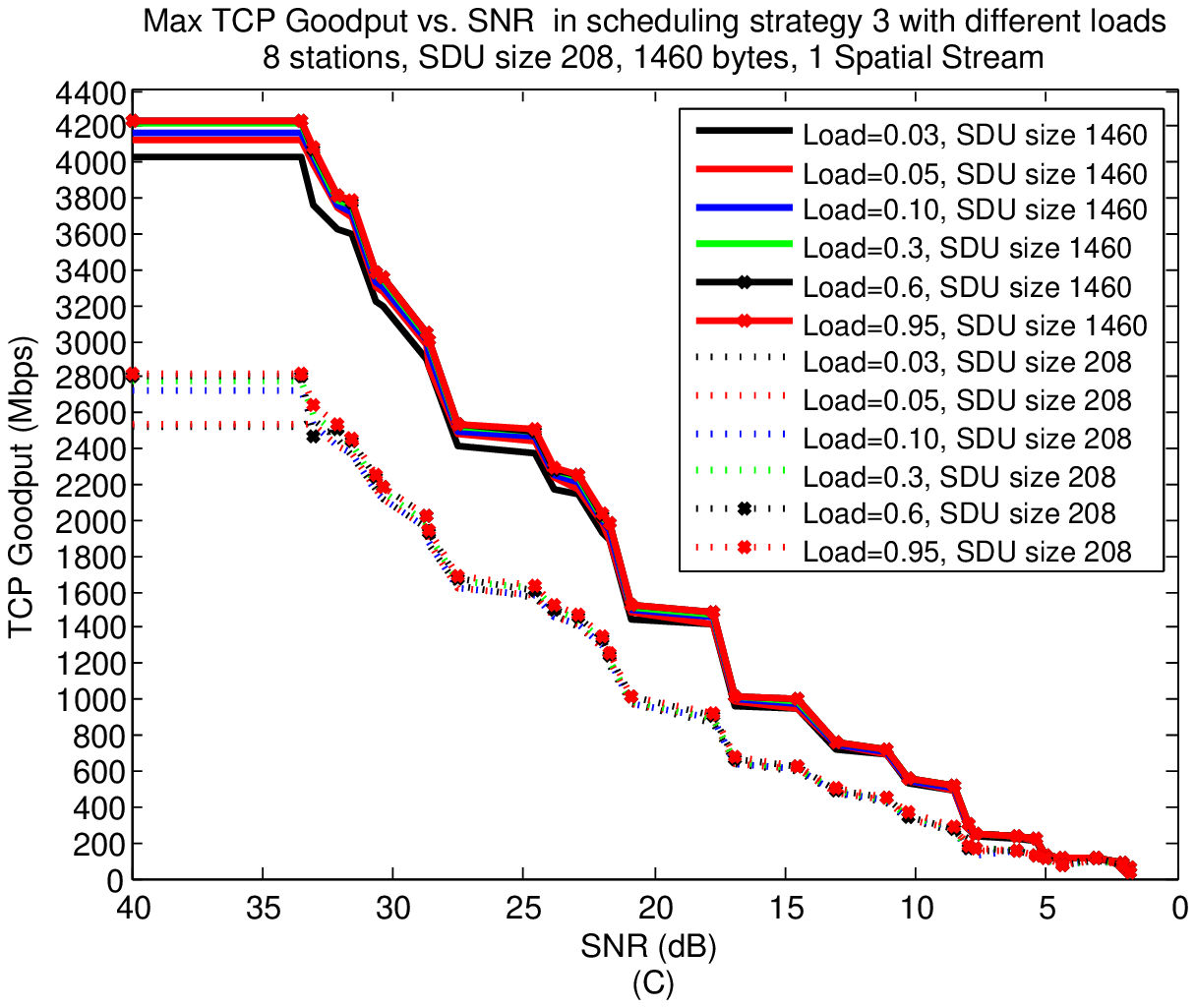}
\includegraphics{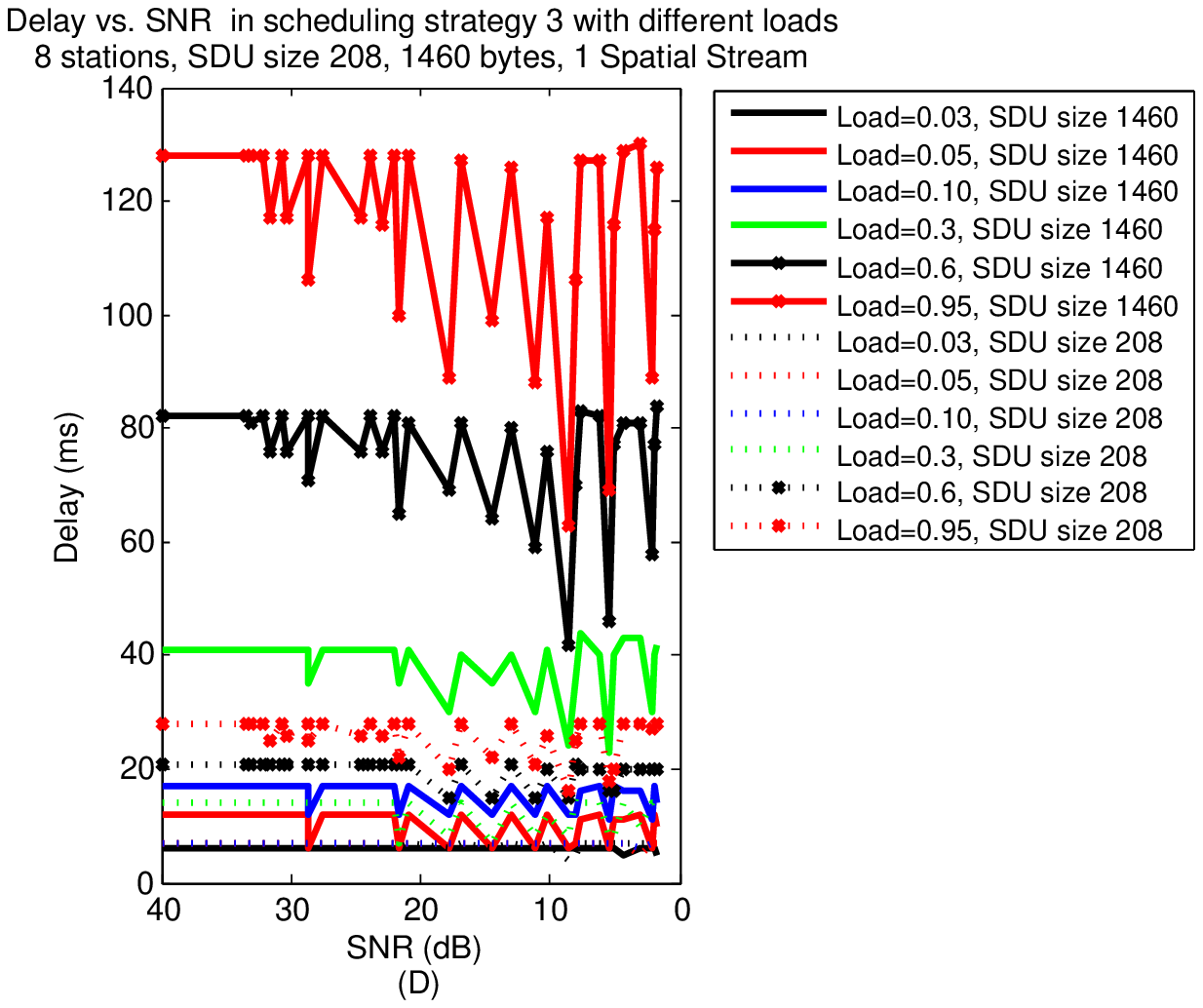}
\caption{TCP Goodput and Delay results for 8 stations and TCP Data segments of 208 and 1460 bytes, for the three scheduling strategies.}
\label{fig:80thrdelsnrc}
\end{figure}

\begin{figure}
\vskip 16cm
\includegraphics{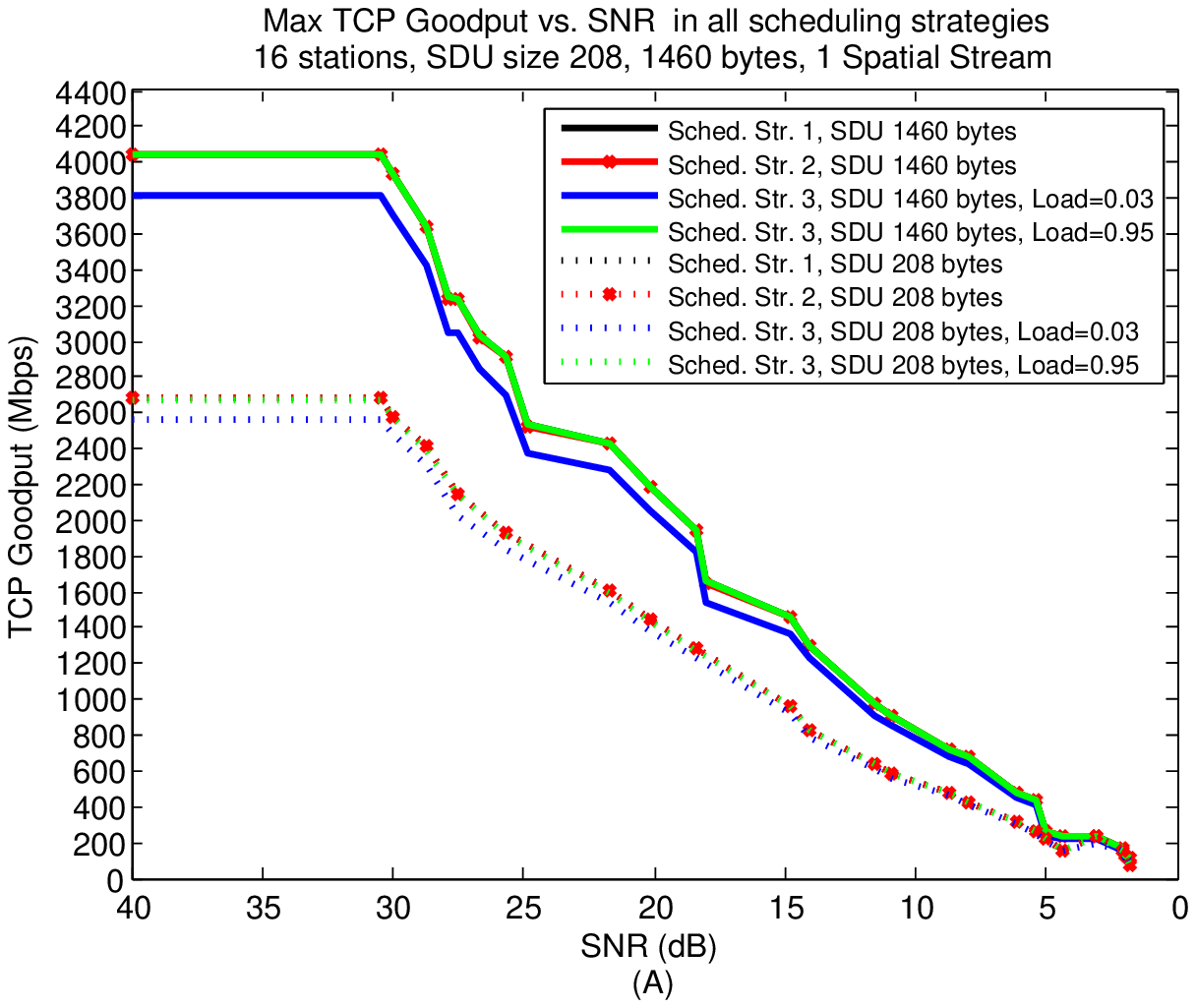}
\includegraphics{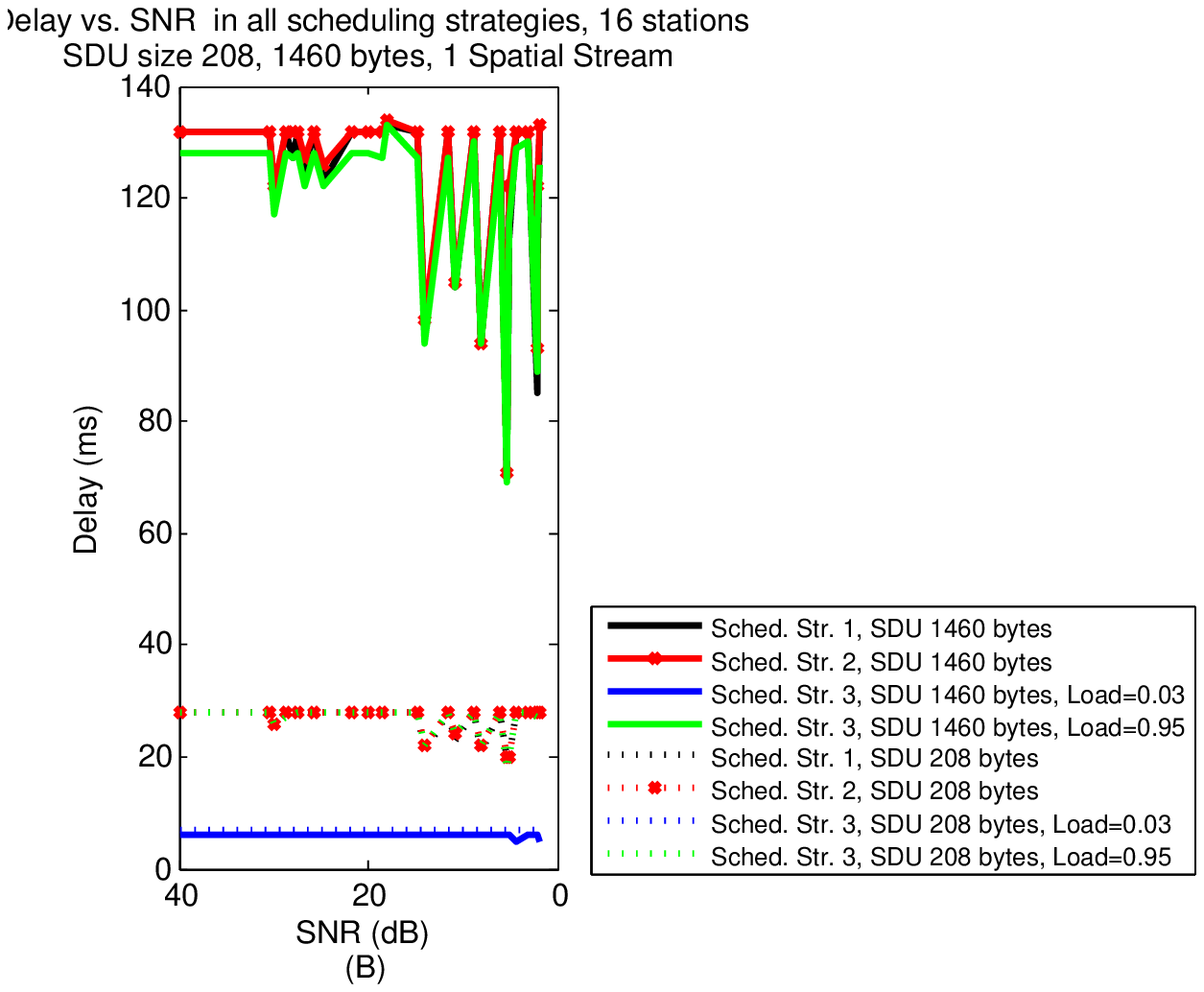}
\includegraphics{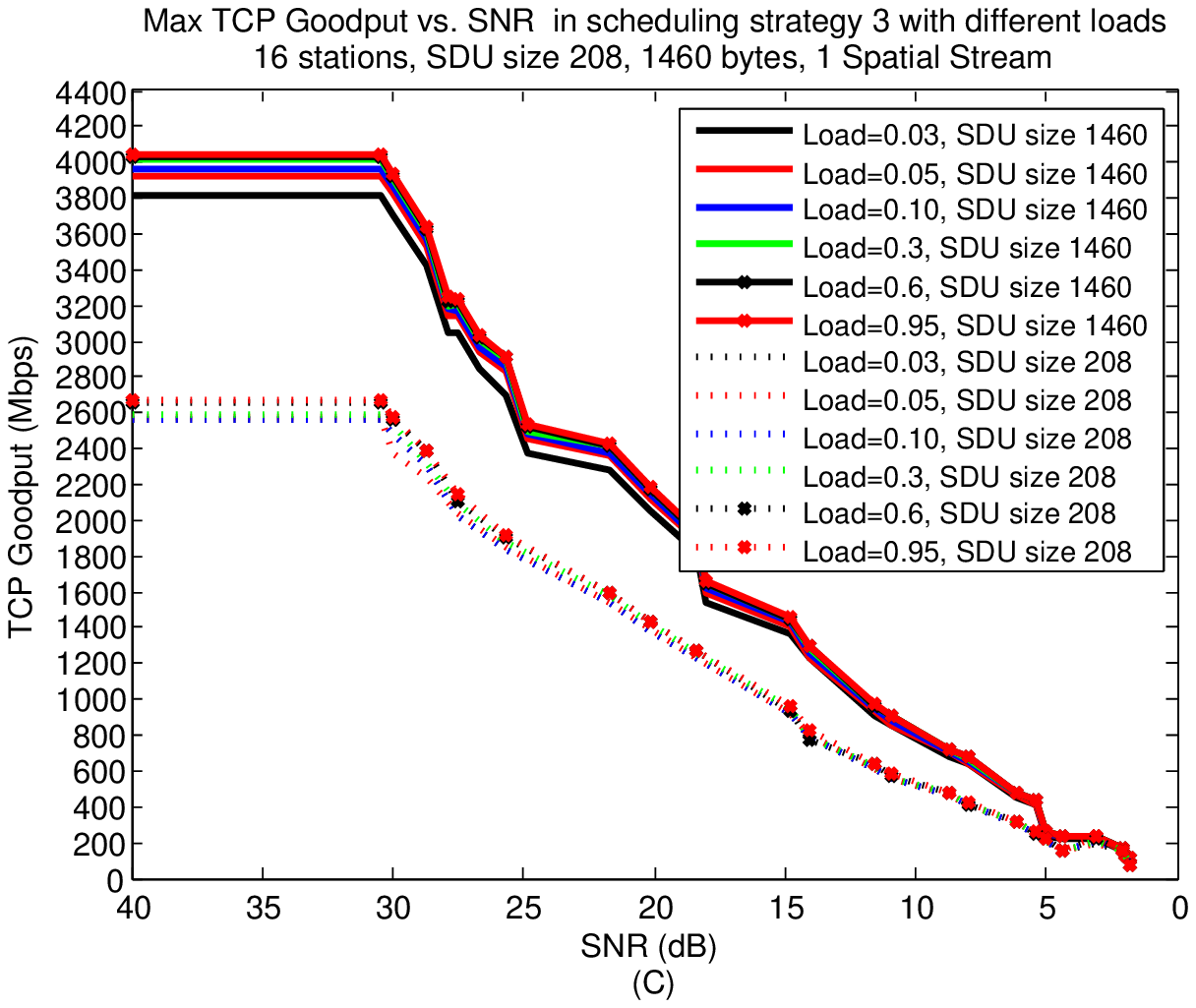}
\includegraphics{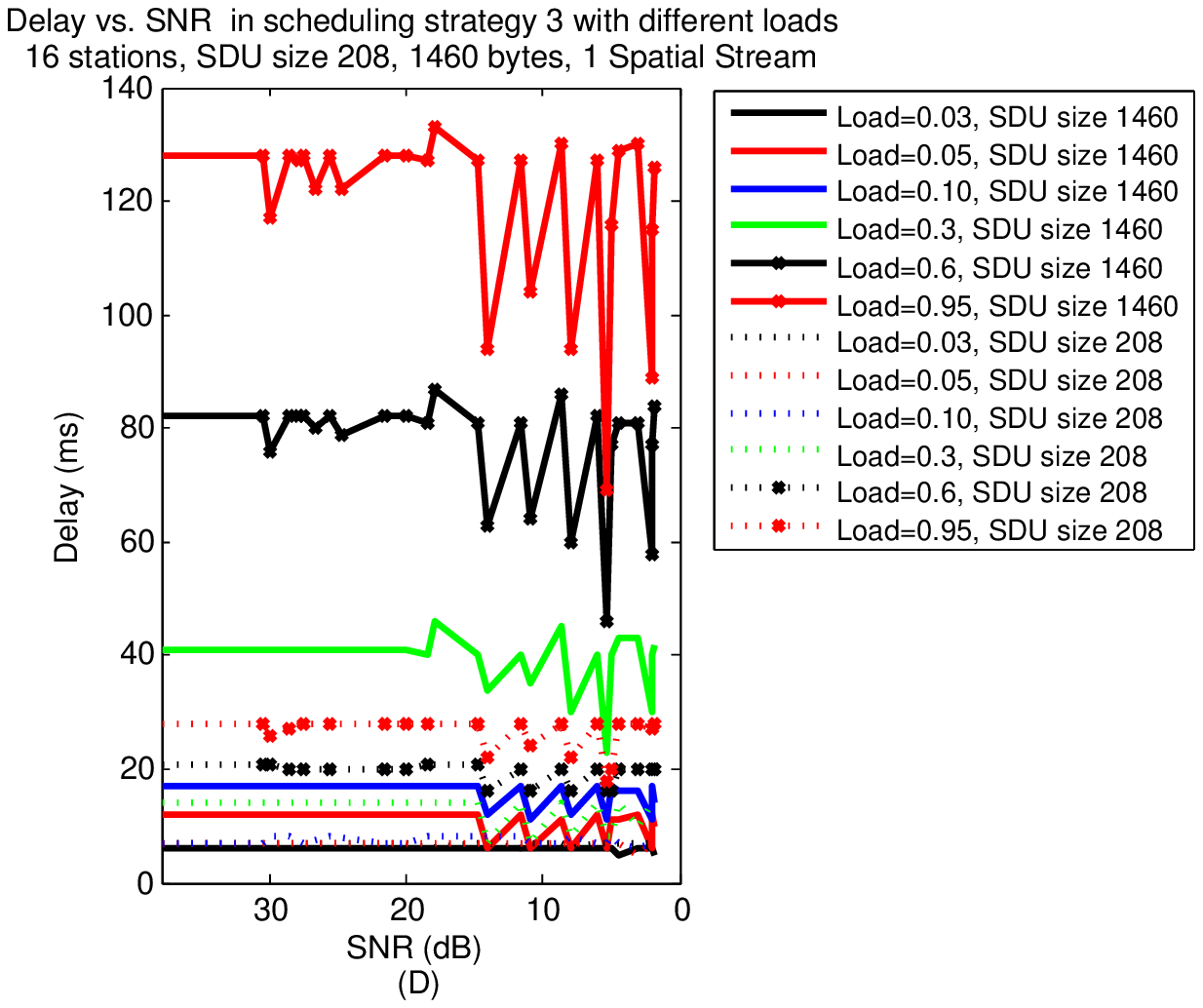}
\caption{TCP Goodput and Delay results for 16 stations and TCP Data segments of 208 and 1460 bytes, for the three scheduling strategies.}
\label{fig:40thrdelsnrc}
\end{figure}

\begin{figure}
\vskip 16cm
\includegraphics{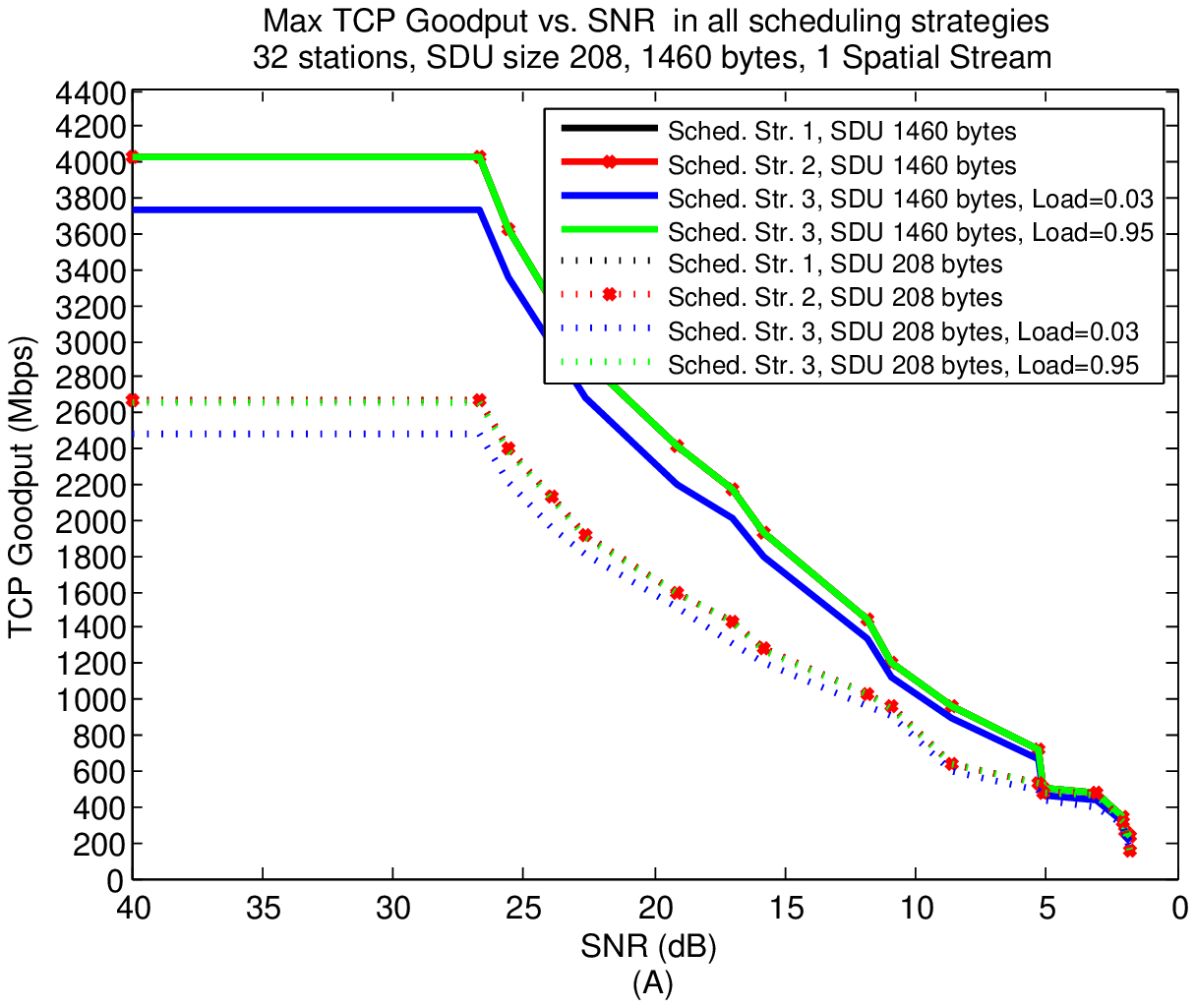}
\includegraphics{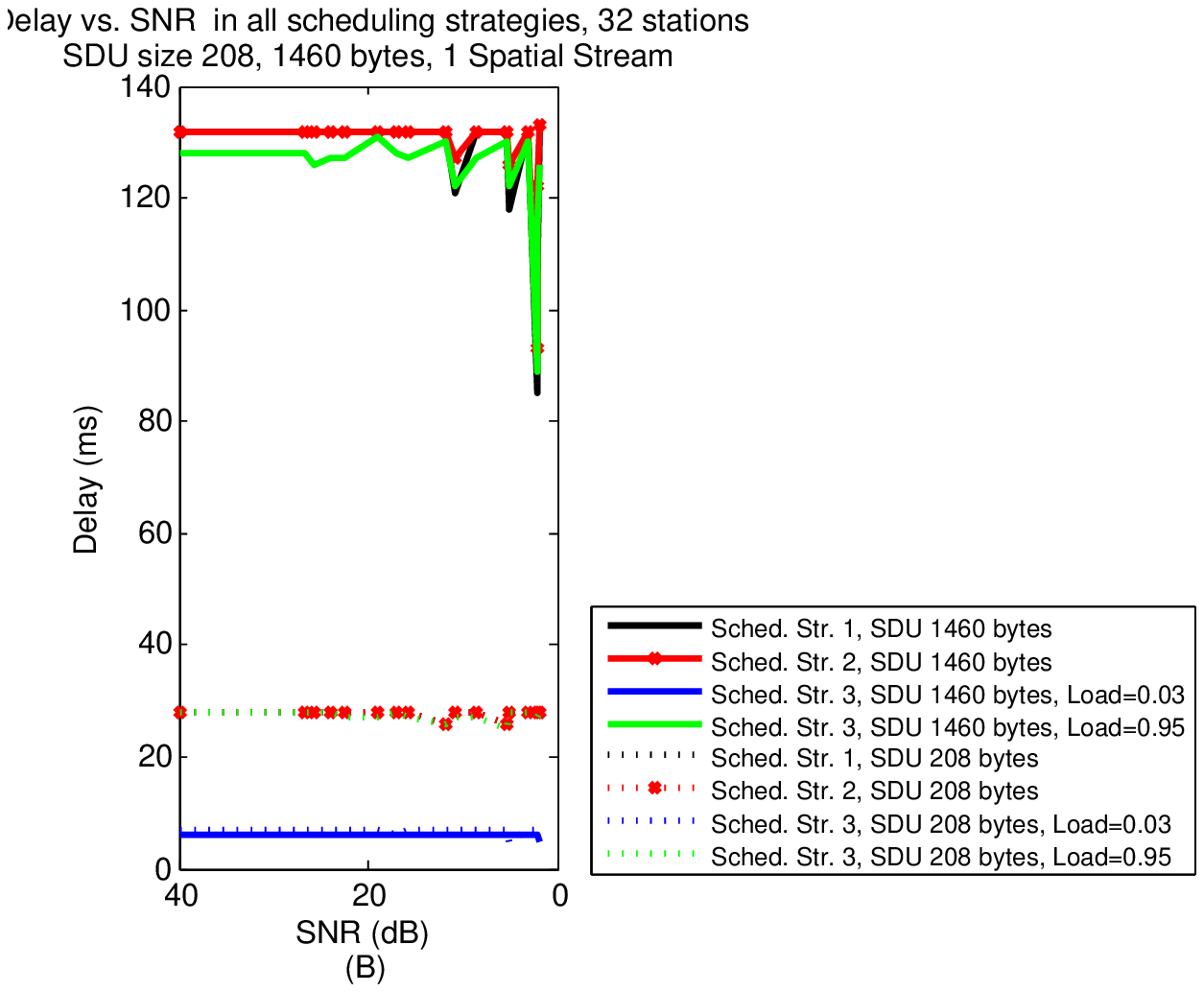}
\includegraphics{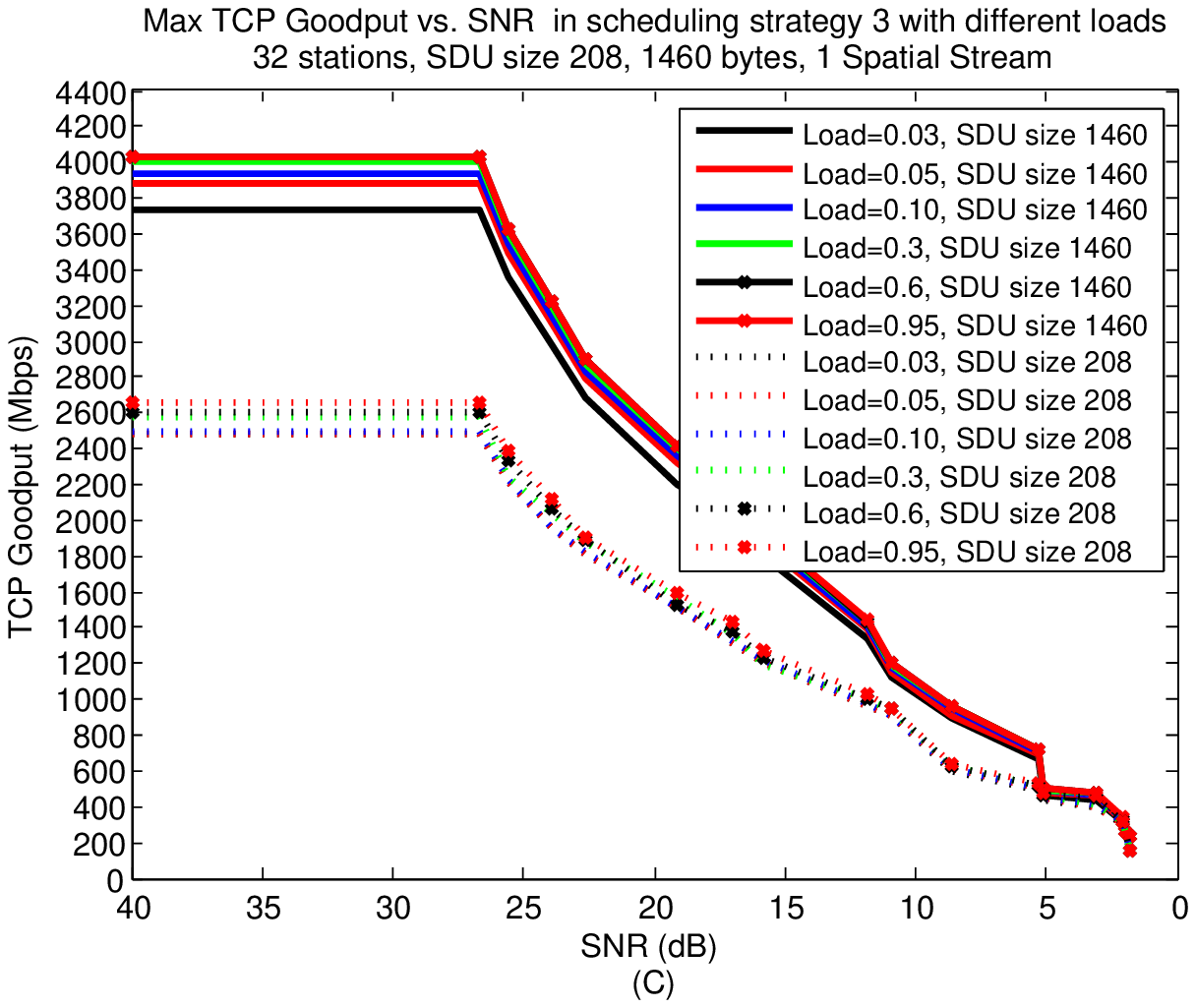}
\includegraphics{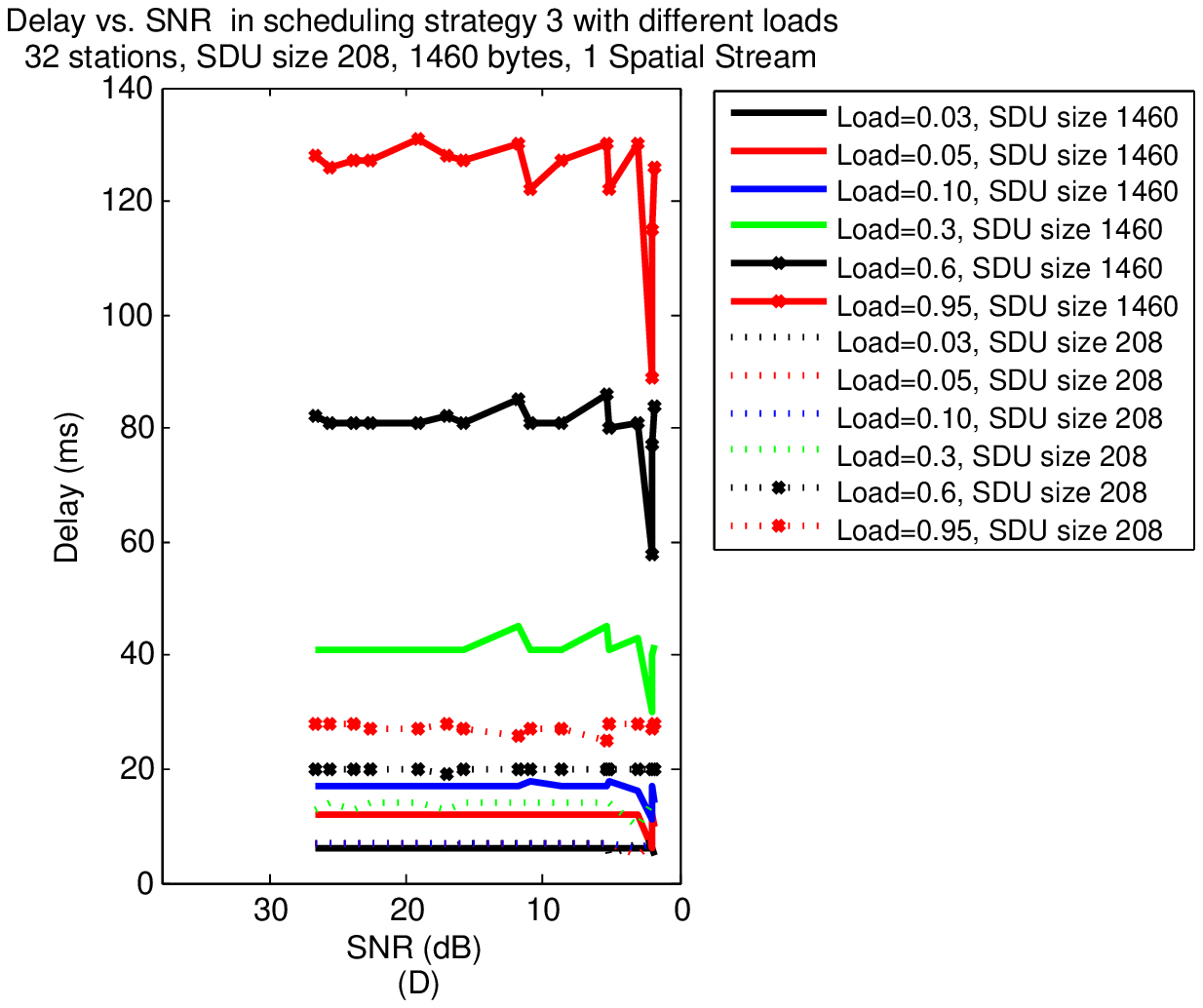}
\caption{TCP Goodput and Delay results for 32 stations and TCP Data segments of 208 and 1460 bytes, for the three scheduling strategies.}
\label{fig:20thrdelsnrc}
\end{figure}

In Figures~\ref{fig:80thrdelsnrc}-\ref{fig:20thrdelsnrc}
we show the same results as in Figure~\ref{fig:160thrdelsnrc}
for 8, 16 and 32 stations, i.e. 80, 40 and 20MHz RUs respectively.
The relative results among the scheduling strategies remain
the same as in Figure~\ref{fig:160thrdelsnrc}. 
Worth mentioning is that
the RU allocations of 160 and 80 MHz have a slightly larger maximum TCP 
Goodput than in RU allocations of 40 and 20 MHz. This is explained by
the PHY rates shown in Table~\ref{tab:phyrates}.
A RU allocation of 80 MHz has half the PHY rate of that of a RU of
160 MHz and this is also the case for RUs of 20 and 40 MHz
respectively. However, a RU of 40 MHz has less than
half the PHY rate than that of a RU of 80 MHz.

In Figure~\ref{fig:load003snr}
we show the TCP Goodput and Delay results for scheduling
strategy 3 with load$=$0.03 only, as a function of the
SNR. Results are shown for 4, 8, 16 and 32 stations,
and for both sizes of TCP Data segments, 
to further check the impact of the number
of stations on the TCP Goodput and Delay.
As was observed before in RUs of 40 and 20 MHz there is
a slightly lower TCP Goodput than in RUs of 160 and 80 MHz.
Concerning the Delay metric, the Delays in the case
of 208 and 1460 bytes TCP Data segments are the same over
all the channels' bandwidths. 

Notice the Delay metrics in the cases
of TCP Data segments of 1460 and 208 bytes in
Figures~\ref{fig:load003snr}(B) and~\ref{fig:load003snr}(D) respectively.
The delay in the case of TCP Data segments of 1460
bytes is smaller than that of 208 bytes: $6.5 ms$ vs.
$7.0 ms$. One might think that as the TCP Data segments
are longer, so are the TXOPs and the Delay metric.
However, in load$=$0.03 only one PPDU is needed for
the transmission of all the TCP Data segments.
Since we assume that the AP transmits
the minimum between the load and the number
of TCP Data segments  possible in one PPDU,
the AP transmits more TCP Data segments 
of 208 bytes than TCP Data segments of 1460 bytes.  
Thus, more TCP Acks
are transmitted in one TXOP in the
case of TCP Data segments of 208 bytes, and so the TXOPs in
this case are also slightly
longer.

\begin{figure}
\vskip 16cm
\includegraphics{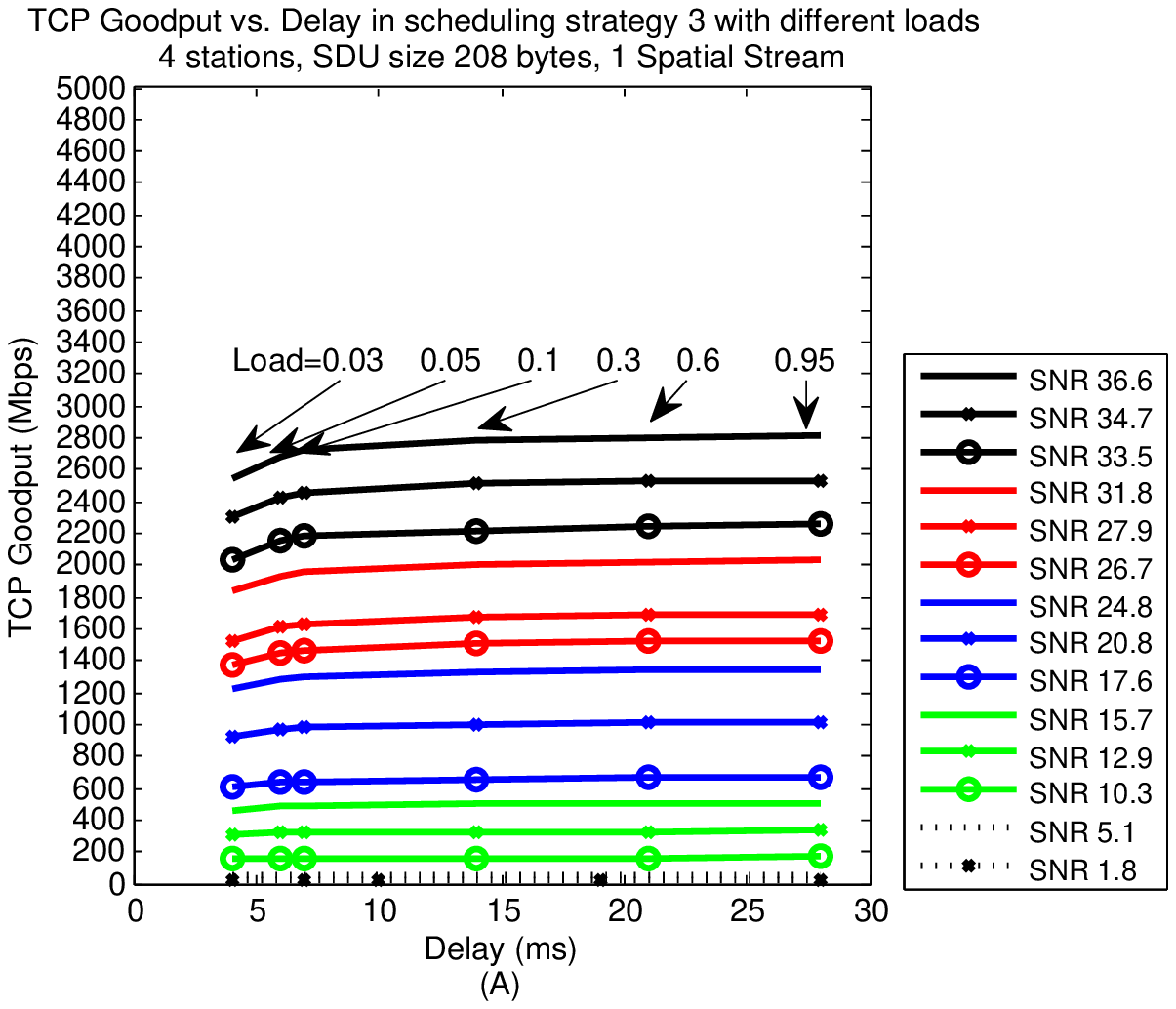}
\includegraphics{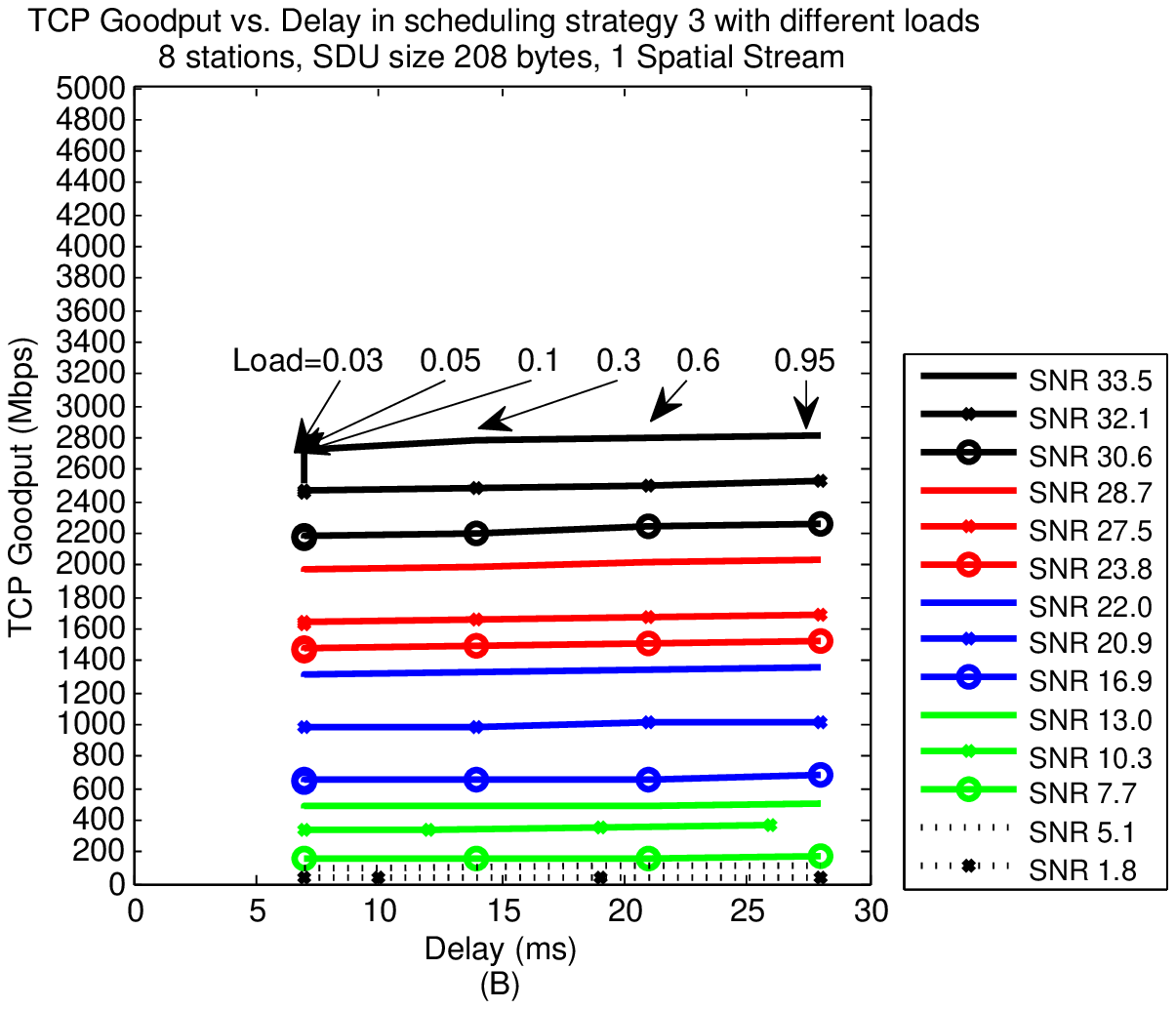}
\includegraphics{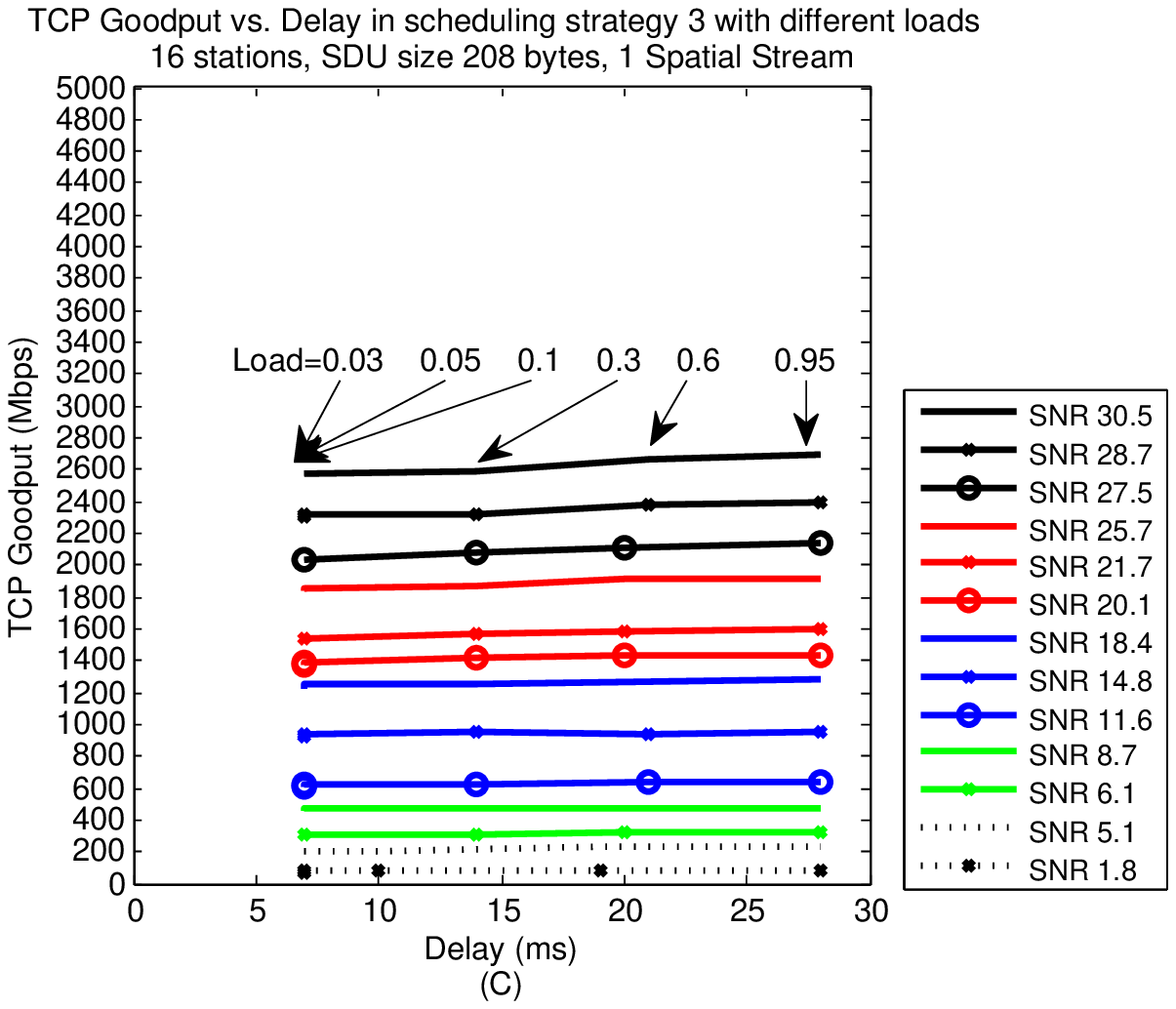}
\includegraphics{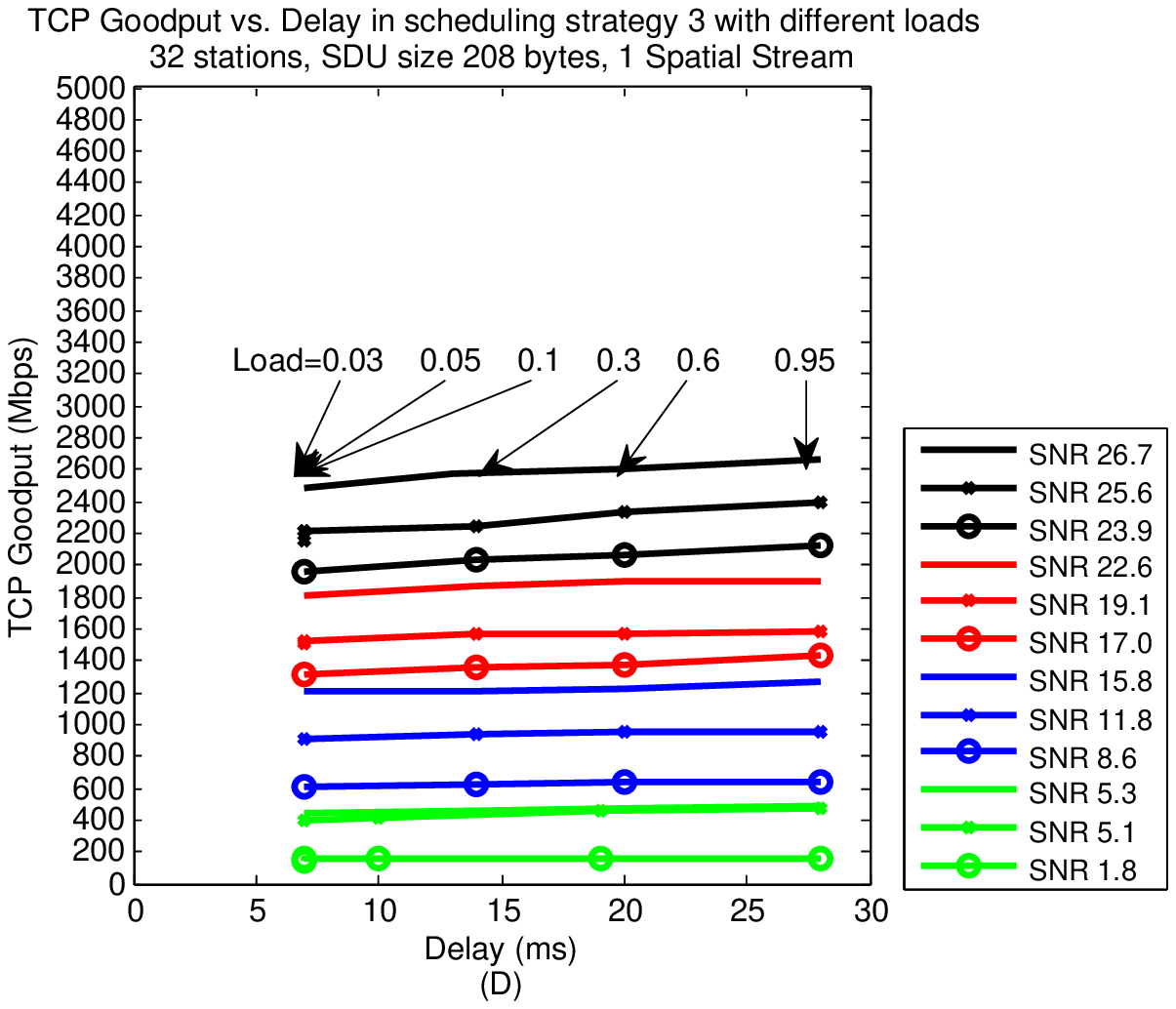}
\caption{TCP Goodput vs. Delay in scheduling strategy 3 with different loads and in different SNRs (MCSs). TCP Data segments of 1460 bytes.}
\label{fig:mcs1460}
\end{figure}

\begin{figure}
\vskip 16cm
\includegraphics{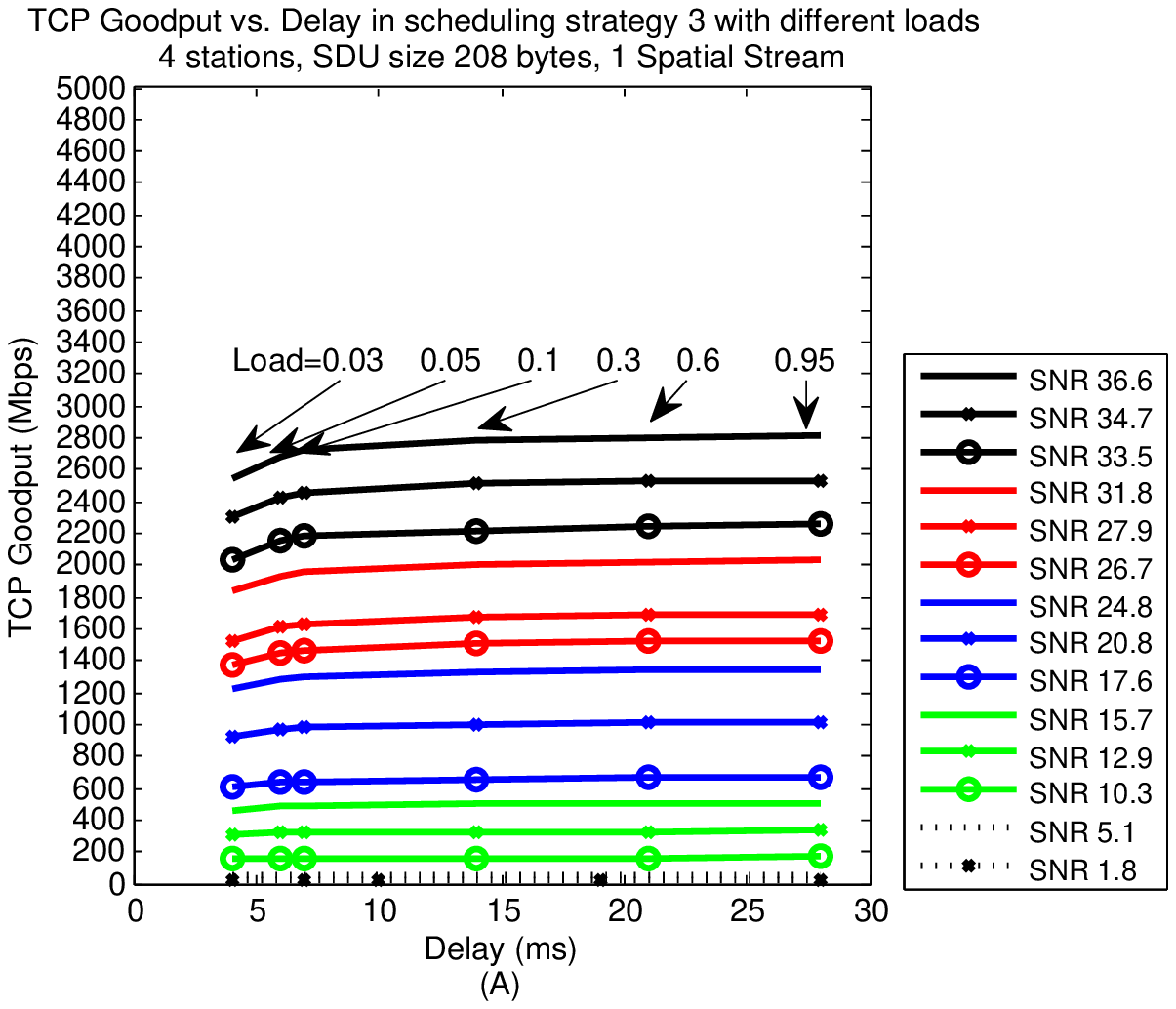}
\includegraphics{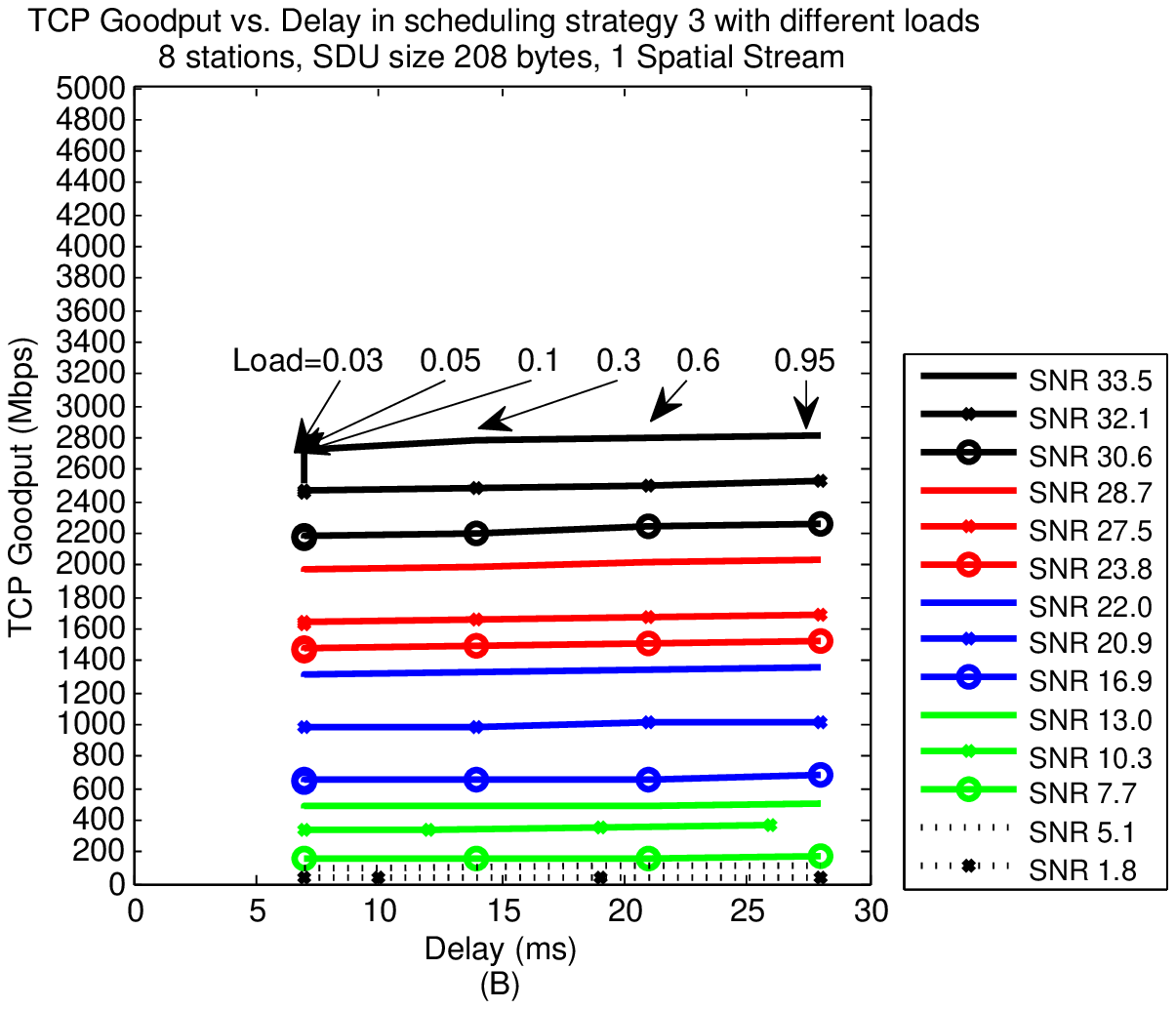}
\includegraphics{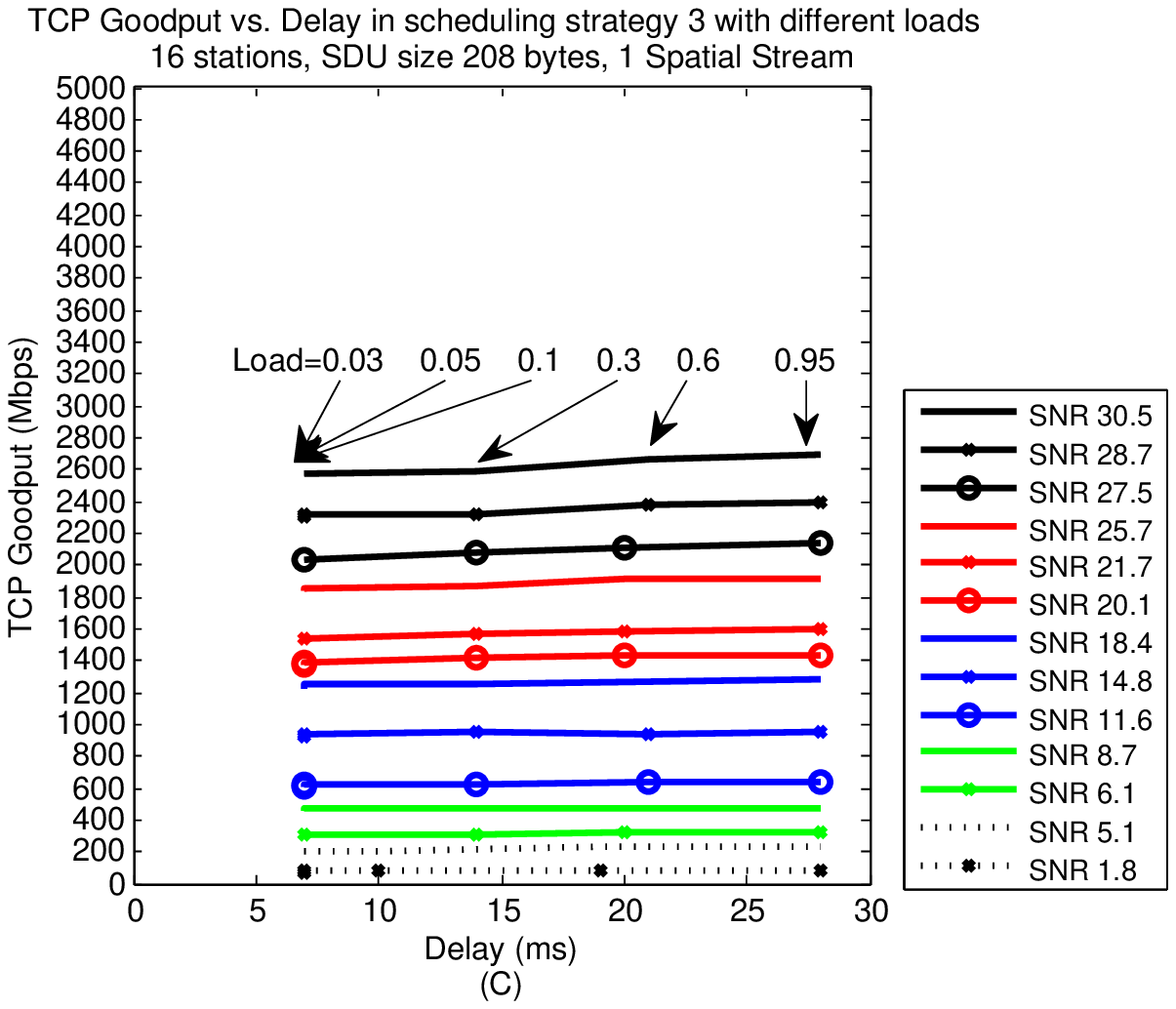}
\includegraphics{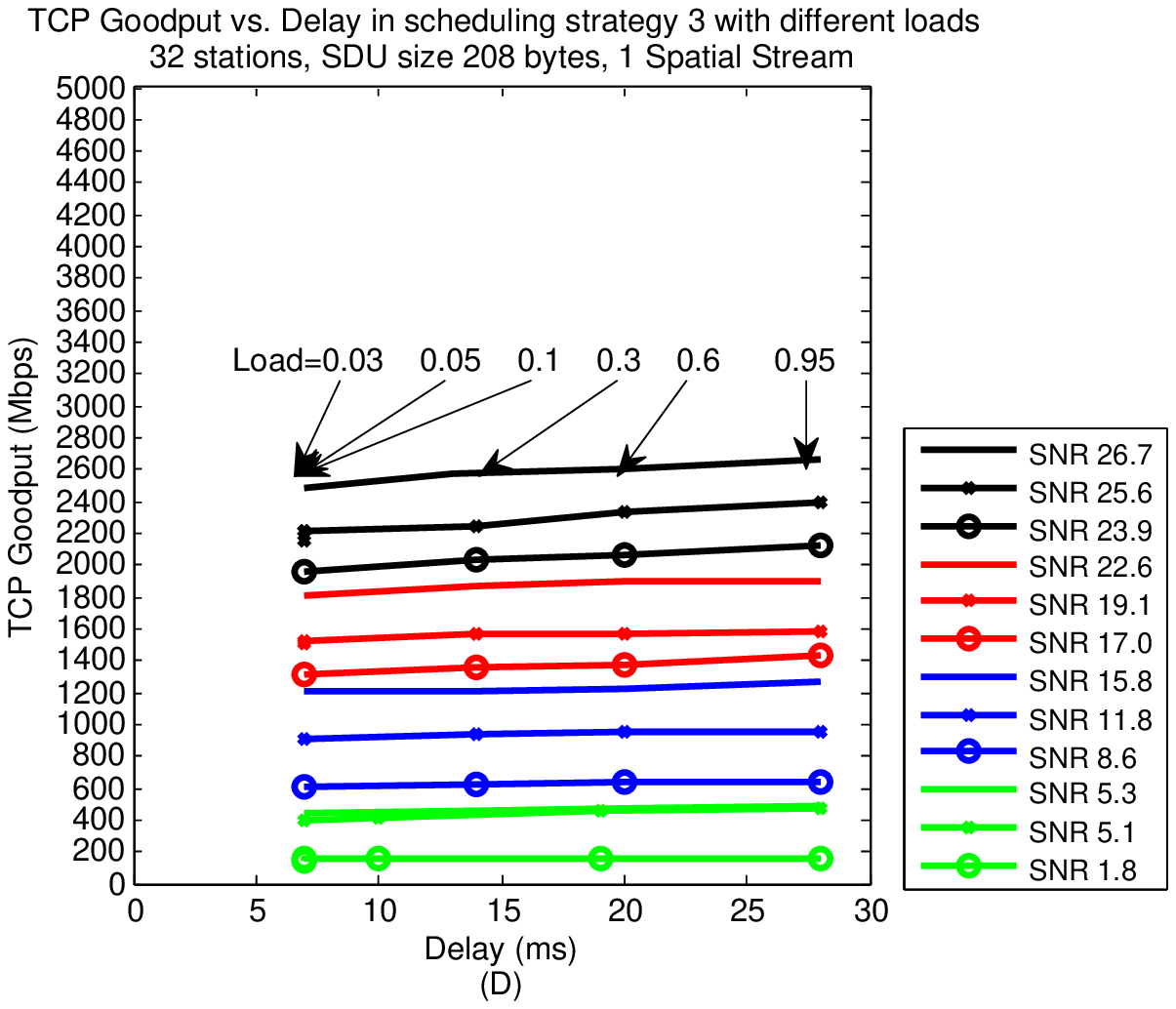}
\caption{TCP Goodput vs. Delay in scheduling strategy 3 with different loads and in different SNRs (MCSs). TCP Data segments of 208 bytes.}
\label{fig:mcs208}
\end{figure}

In Figures~\ref{fig:mcs1460} and~\ref{fig:mcs208} 
we show the impact of the Delay, i.e. length of the TXOPs
on the received TCP Goodput in different SNRs or MCSs in use.
Figures~\ref{fig:mcs1460} and~\ref{fig:mcs208} correspond to TCP 
Data segments of 1460 and 208 bytes respectively.
The Delay metric can be a parameter in scheduling strategy 3
where the load determines the TXOPs length.
In Figure~\ref{fig:mcs1460}(A), (B), (C) and (D)
results are shown for RU allocation for 4, 8, 16 and 32 stations
respectively. One can see that the maximum TCP Goodput
is already received around Delays of 25ms. Thus,
for overloaded TCP connections and long TCP Data
segments RTTs of 25ms are sufficient to use the
11ax channel efficiently. For TCP Data segments
of 208 bytes the sufficient Delay is about 13ms, Figure~\ref{fig:mcs208}.

\begin{figure}
\vskip 3cm
\includegraphics{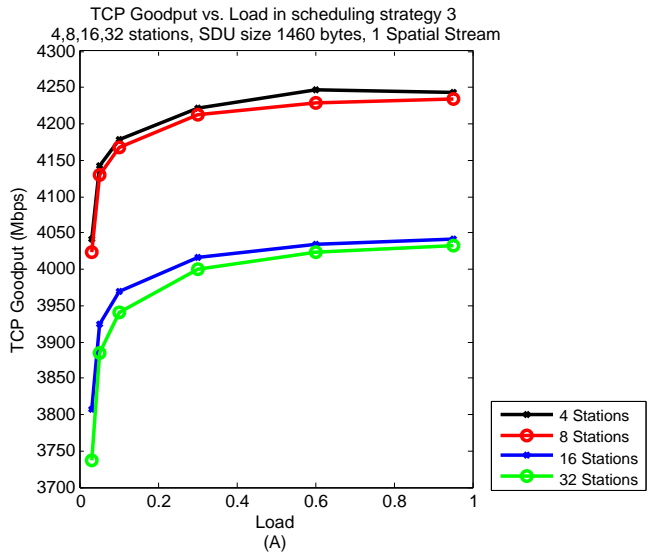}
\includegraphics{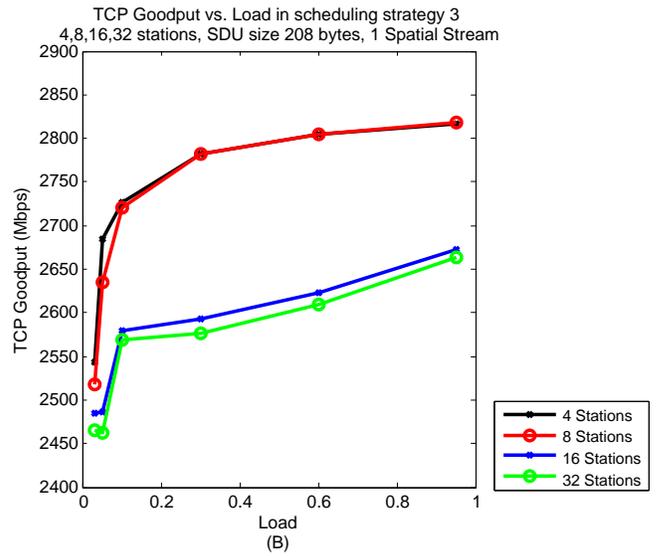}
% \special{psfile= del40thr.ps voffset=-165 hoffset= -100 hscale=70 vscale=70}
% \special{psfile= del20thr.ps voffset=-165 hoffset= 160 hscale=70 vscale=70}
\caption{ Maximum TCP Goodput vs. load in scheduling strategy 3 for 4, 8, 16 and 32 stations and for TCP Data segments of 208 and 1460 bytes.}
\label{fig:thrload}
\end{figure}

Finally, in Figure~\ref{fig:thrload}
we show the TCP Goodput vs. load in scheduling strategy 3
for 4, 8, 16 and 32 stations
and for TCP Data segments of 208 and 1460 bytes.
For TCP Data segments
of 1460 bytes and when using load$=$0.03
for all the number of stations, one only loses
about $5.0 \%$ of the maximum Goodput .
When TCP Data segments are of 208 bytes one loses
about $8.0 \%$. The difference between $5.0 \%$
to $8.0 \%$ is due to the fact that
when short TCP Data segments are transmitted, more
of them are needed to receive a good relation between
the TCP Data bits transmitted and the overhead.

\newpage

\newpage

\newpage

% \newpage

% \begin{figure}
% \vskip 16cm
% % \special{psfile= ../simul/all/thrload208.ps voffset=90 hoffset= -100 hscale=70 vscale=70}
% % \special{psfile= ../simul/80mhz/res/del80thr.ps voffset=90 hoffset= 160 hscale=70 vscale=70}
% % \special{psfile= ../simul/40mhz/res/del40thr.ps voffset=-165 hoffset= -100 hscale=70 vscale=70}
% % \special{psfile= ../simul/20mhz/res/del20thr.ps voffset=-165 hoffset= 160 hscale=70 vscale=70}
% \caption{}
% \label{fig:res151}
% \end{figure}

% \input res
% \input thrres
\clearpage

\section{Summary}

In this paper we suggest TCP-aware scheduling strategies
for the transmission of unidirectional TCP traffic
between the AP and the stations in an IEEE 802.11ax
BSS. The AP is the source of TCP Data segments and the stations
reply with TCP Acks. Our proposed TCP-aware
scheduling strategies take
advantage of MU transmission capability
over the Uplink, which is a feature first introduced
in IEEE 802.11ax .

Our TCP-aware scheduling strategies are based on TXOP; 
the selected MCSs used over
the Uplink ensure reliable transmissions
while the Downlink is unreliable. We measure the
Goodput and TXOPs' length in different transmission
strategies that determine the TXOP formation
that controls how many TCP Data
segments are transmitted in a cycle.

Our main finding is that relatively short
TXOPs cycles are sufficient to receive almost
the largest Goodput, and there is a parameter
for TXOP formation
by which one can control what Goodput
to receive, while paying in the TXOP cycle length.
The TXOP cycle length is important since it is a part
of the time-out measured by TCP.
For example, we checked the Goodput and cycles' length
for 1460 and 208 bytes TCP Data segments' length.
In order to receive the largest Goodput
one needs cycles of $25ms$ and $13ms$ respectively.
If one loses $5-8 \%$ of the maximum Goodput,
the cycles' length can decrease to $3-5 ms$.

Further research issues concerning TCP traffic
over IEEE 802.11ax can be to explore
the transmission of TCP data traffic
from the stations to the AP and transmitting
bidirectional TCP data
traffic. Also, a small transmission error
probability over the Uplink
can be achieved 
by using large indexed MCSs which enable large PHY rates,
while on the other hand using shorter MPDUs and so transmitting
a smaller number of TCP Acks.
It is interesting to check if such a scheme
is more efficient than the one used in this paper
where lower indexed MCSs ensure a reliable
Uplink with lower PHY rates.

\newpage
\clearpage

\section{Appendix}

In this appendix we demonstrate the relation
between the channel bandwidth, MCS in use and
the SNR to the received BER
in various RUs' bandwidths and
MCSs in use. The results are taken
from~\cite{IEEEber}. We assume
that the AP is communicating with every station
over one spatial stream.

In Table~\ref{tab:snrber} we show some examples for this
relation, assuming the 160MHz channel RU allocation. For example, for
$SNR=36.6 dB$ the BER equals 0 for all the MCSs and 
clearly in this case both the AP and the stations use
MCS11 over the DL and the UL respectively,
which enables the largest PHY rate. For $SNR=35.1$
one can assume that either the AP uses MCS11 and the
stations use MCS10, or both the AP and the stations
use MCS10. Recall that in this paper we assume that
the channel over which the TCP Acks are transmitted,
i.e. the UL,
is reliable, and so 
in SNR$=$35.1
the stations cannot use MCS11
over the UL.
Notice that the AP can use either MCS11 or MCS10
over the DL, with a trade-off: the first possibility
is a larger PHY rate but BER$>$0, while the
other possibility is the PHY rate is smaller but BER$=$0.

In Table~\ref{tab:snrber} we also show
the most efficient MCSs to use for every SNR (in terms of the Goodput)
over the DL and
UL. These MCSs are shown in {\it Italic}. 
For example, for SNR$=$36.6 dB MCS11 is used over the DL
and the UL. For SNR$=$35.1 dB MCS11 is used over the DL
and MCS10 is used over the UL.
Notice also that
when considering a 160MHz channel RU allocation, such a channel
can be used for SNRs larger than 10.2 dB only.

\begin{table}
\caption{\label{tab:snrber}{The relation between the SNR, MCS and BER
in a 160 MHz channel and 1 Spatial Stream}}
\vspace{3 mm}
\tiny
\center
\begin{tabular}{|c|c|c|c|c|c|c|c|c|c|c|c|c|}  \hline
 SNR    & \multicolumn{12}{|c|}{MCS} \\  \cline{2-13}
  & 0 &1 &2 &3 &4 &5 &6 &7 &8 &9 &10 &11  \\ \hline
 36.6   & 0 & 0 & 0 & 0 & 0 & 0 & 0 & 0 & 0 & 0 & 0 & {\it 0}  \\ 
 35.1   & 0 & 0 & 0 & 0 & 0 & 0 & 0 & 0 & 0 & 0 & {\it 0} & 0.6595  \\ 
 34.0   & 0 & 0 & 0 & 0 & 0 & 0 & 0 & 0 & 0 & {\it 0} & {\it 0.0222} & 1 \\ 
 33.6   & 0 & 0 & 0 & 0 & 0 & 0 & 0 & 0 & 0 & {\it 0} & 0.3696 & 1 \\ 
 33.5   & 0 & 0 & 0 & 0 & 0 & 0 & 0 & 0 & {\it 0} & {\it 0.0005} & 0.4958 & 1 \\ 
 32.5   & 0 & 0 & 0 & 0 & 0 & 0 & 0 & 0 & {\it 0} & {\it 0.0987} & 1 & 1  \\ 
 31.7   & 0 & 0 & 0 & 0 & 0 & 0 & 0 & 0 & {\it 0} & 0.9412 & 1 & 1  \\ 
 30.2   & 0 & 0 & 0 & 0 & 0 & 0 & 0 & {\it 0} & 0.6457 & 1 & 1 & 1  \\ 
 27.1   & 0 & 0 & 0 & 0 & 0 & 0 & {\it 0} & {\it 0.0749} & 1 & 1 & 1 & 1  \\ 
 25.9   & 0 & 0 & 0 & 0 & 0 & {\it 0} & {\it 0.0022} & 1 & 1 & 1 & 1 & 1  \\ 
 24.7   & 0 & 0 & 0 & 0 & {\it 0} & {\it 0.0010} & 0.832 & 1 & 1 & 1 & 1 & 1  \\ 
 20.1   & 0 & 0 & 0 & {\it 0} & {\it 0.1417} & 1 & 1 & 1 & 1 & 1 & 1 & 1  \\ 
 19.7   & 0 & 0 & 0 & {\it 0} & 0.5970 & 1 & 1 & 1 & 1 & 1 & 1 & 1  \\ 
 17.5   & 0 & 0 & {\it 0} & {\it 0.0005} & 1 & 1 & 1 & 1 & 1 & 1 & 1 & 1  \\ 
 16.4   & 0 & 0 & {\it 0} & 0.4973 & 1 & 1 & 1 & 1 & 1 & 1 & 1 & 1  \\ 
 14.6   & 0 & {\it 0} & {\it 0.0003} & 1 & 1 & 1 & 1 & 1 & 1 & 1 & 1 & 1  \\ 
 13.6   & 0 & {\it 0} & 0.4188 & 1 & 1 & 1 & 1 & 1 & 1 & 1 & 1 & 1  \\ 
 10.2   & {\it 0} & 1 & 1 & 1 & 1 & 1 & 1 & 1 & 1 & 1  & 1 & 1  \\ \hline
\end{tabular}  
\end{table}

\begin{figure}
\vskip 16cm
\includegraphics{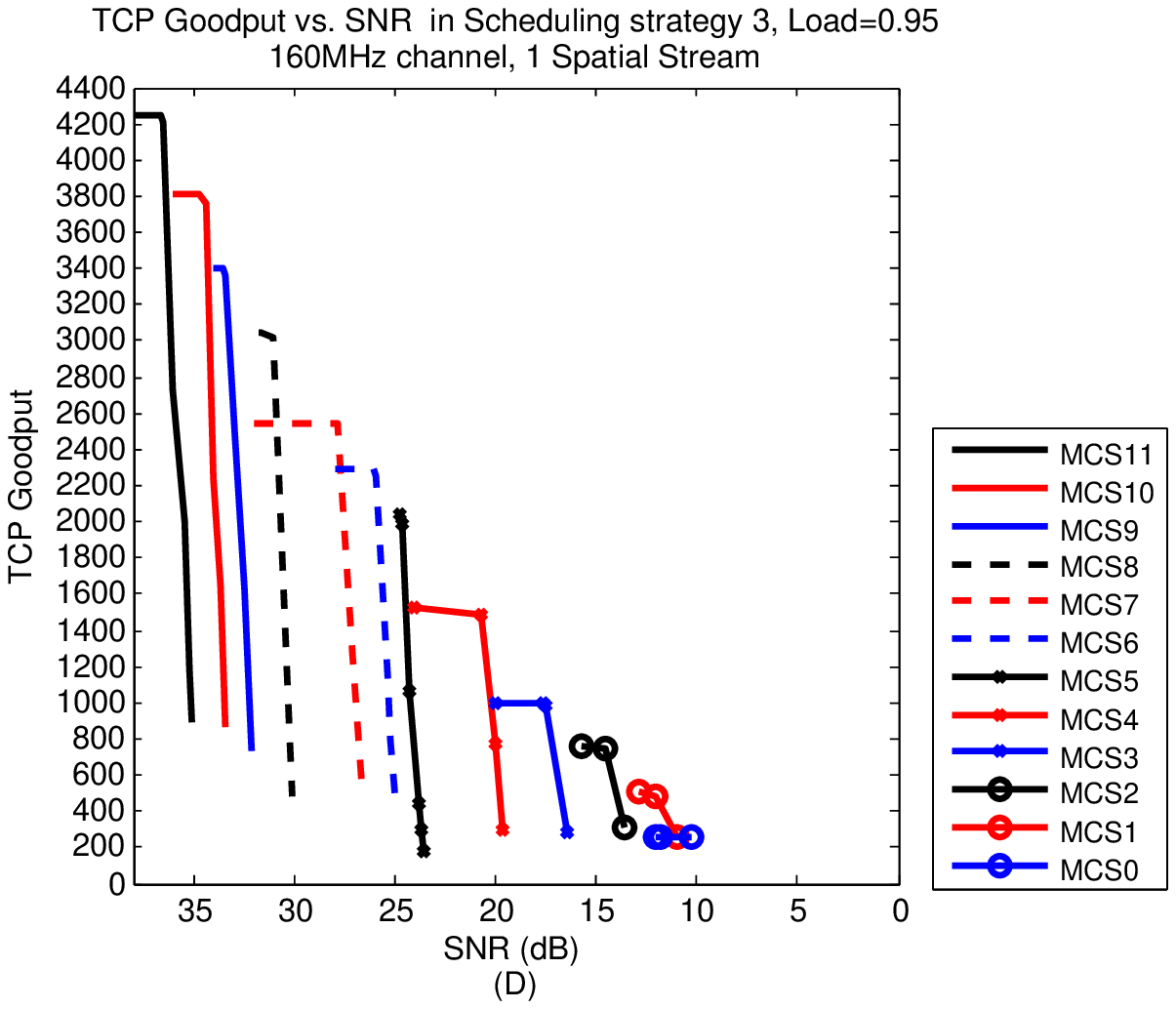}
\includegraphics{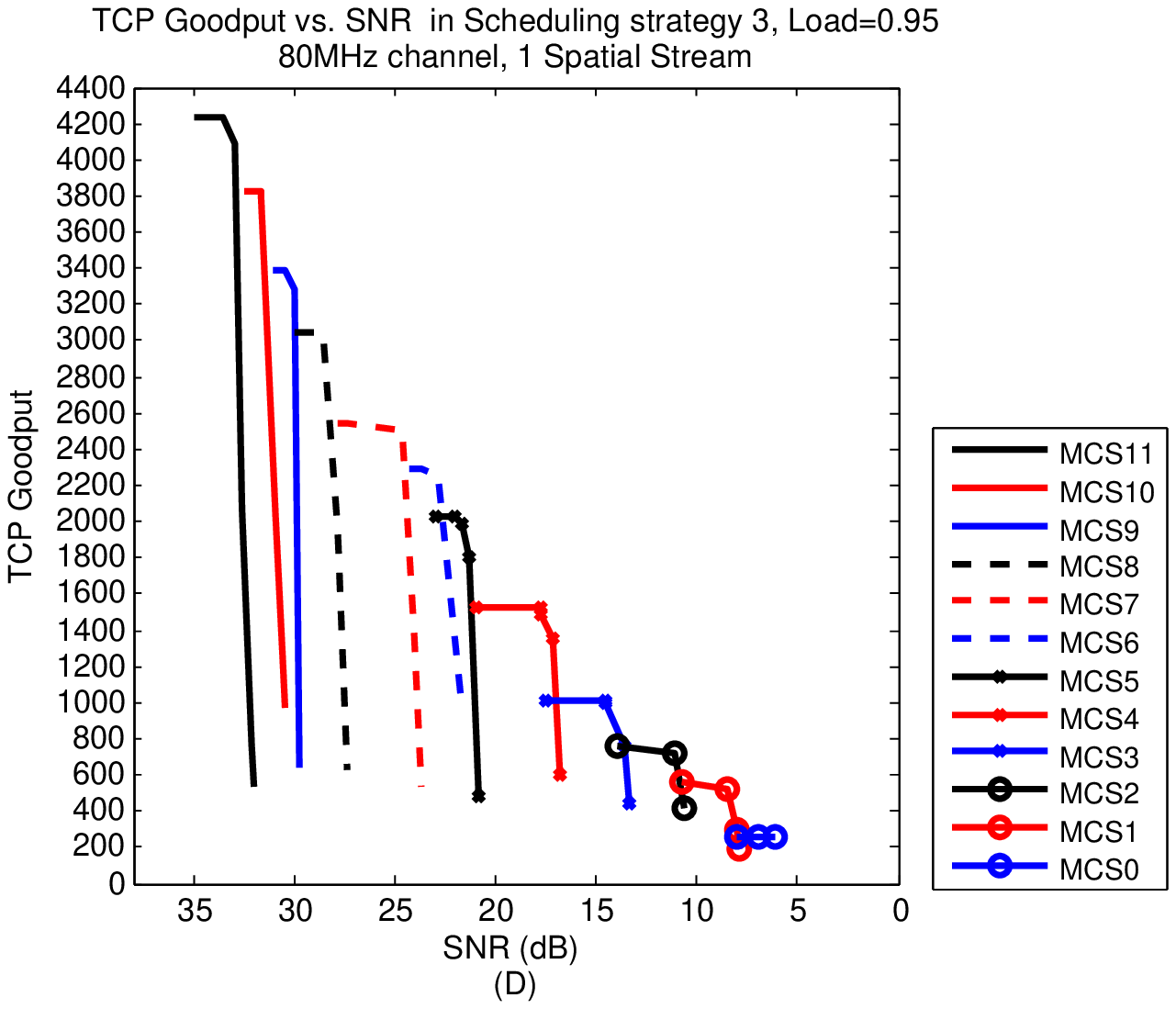}
\includegraphics{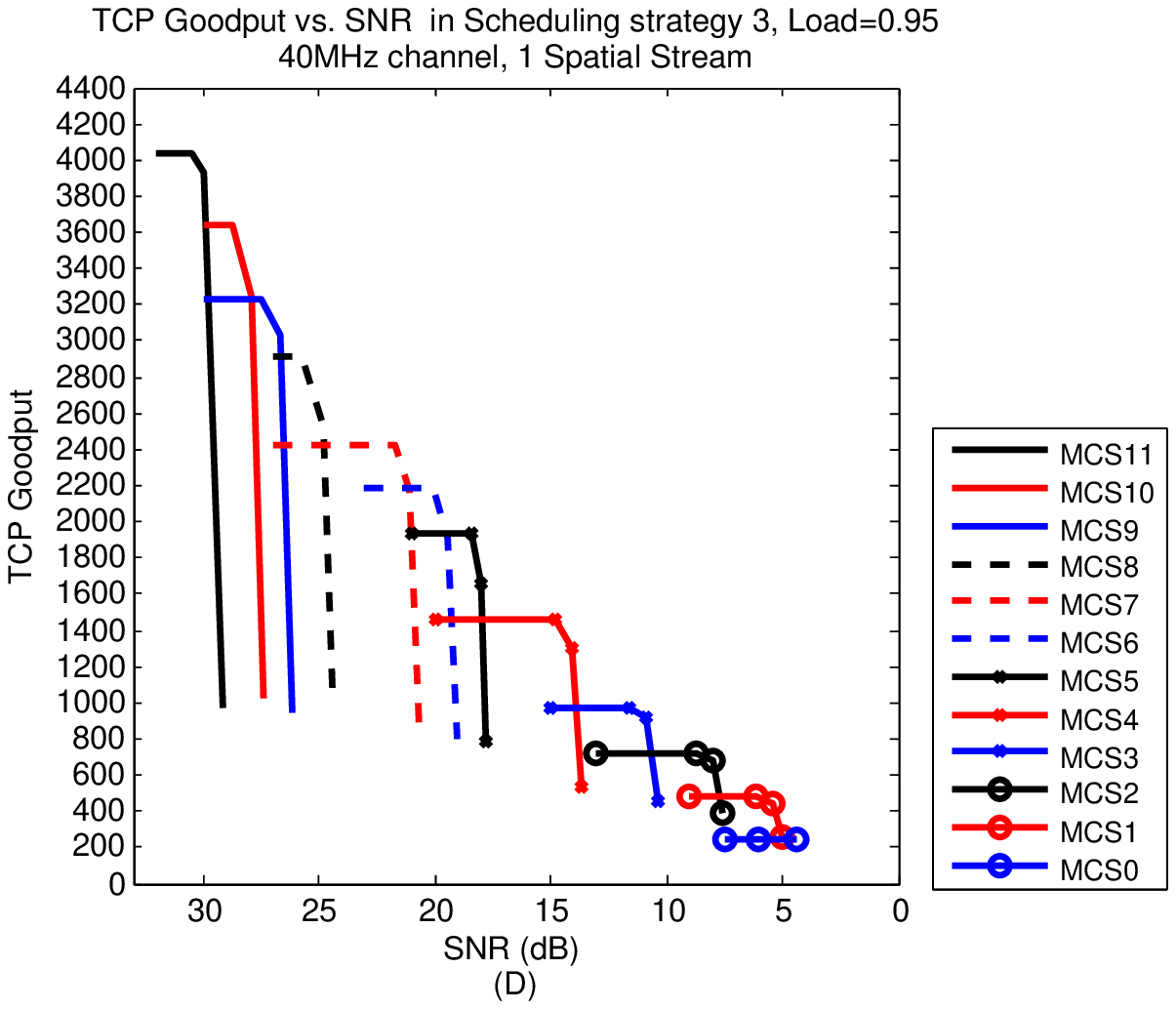}
\includegraphics{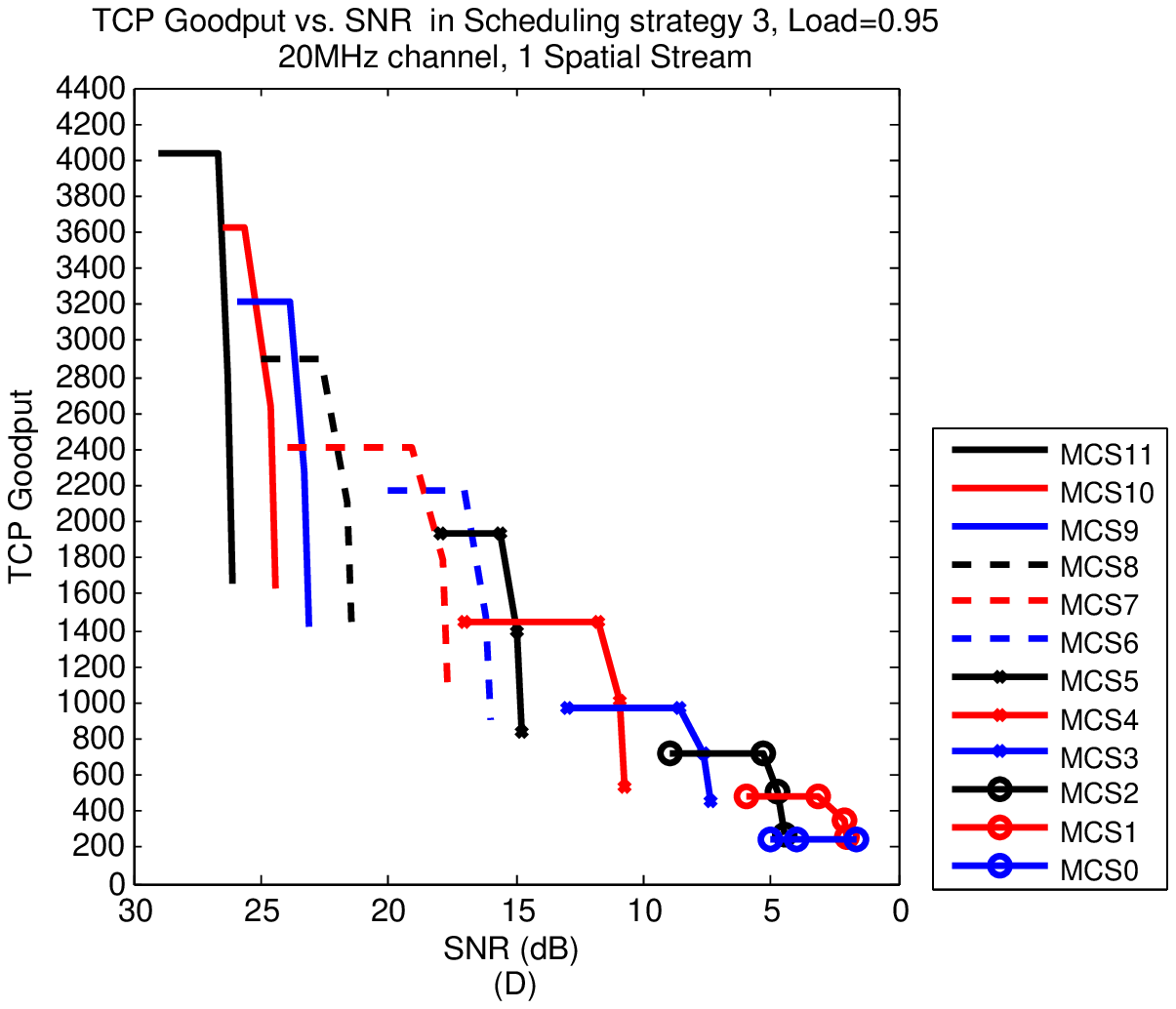}
\caption{The TCP Goodput vs. SNR in various channel bandwidths and MCSs. }
\label{fig:res1}
\end{figure}

In Figure~\ref{fig:res1} we show for the four RU allocation 
channels in use in this
paper, namely 160, 80, 40 and 20 MHz, and for scheduling strategy 3
with $Load=0.95$, the received TCP Goodput as a function of the
SNR. We assume TCP Data segments of
1460 bytes. The curves are generated
based on raw data from~\cite{IEEEber} and
are 'waterfall' shaped.
One can see that every MCS has a range of SNR values
over which it enables  a positive TCP Goodput. The lower
indexed MCSs can be used in lower SNRs, but with a smaller
PHY rate and so with a smaller TCP Goodput.

\clearpage

%%%%%%%%%%%%%%%%%%%%%%%%%%%%%%%%%%%%%%%%%%%%%%%%%%%%%%%%%%

\bibliographystyle{abbrv}
\bibliography{main}

\begin{thebibliography}{10}

\bibitem{IEEEBase1}
\newblock{IEEE Std. 802.11$^{TM}$-2016},
\newblock{IEEE Standard for Information Technology - 
Telecommunications and Information Exchange between Systems - Local
and Metropolitan Area Networks - Specific Requirements. Part 11:
Wireless LAN Medium Access Control (MAC) and Physical Layer (PHY)
Specifications},
\newblock{IEEE, NewYork, (December 2016)}


\bibitem{IEEEax}
\newblock{IEEE P802.11ax$^{TM}$/D1.4},
\newblock{IEEE Draft Standard for Information Technology - 
Telecommunications and Information Exchange between Systems - Local
and Metropolitan Area Networks - Specific Requirements. Part 11:
Wireless LAN Medium Access Control (MAC) and Physical Layer (PHY)
Specific requirements. }
\newblock{IEEE, NewYork, (2017)}

\bibitem{IEEEac}
\newblock{IEEE Std. 802.11ac$^{TM}$-2013},
\newblock{IEEE Standard for Information Technology - 
Telecommunications and Information Exchange between Systems - Local
and Metropolitan Area Networks - Specific Requirements. Part 11:
Wireless LAN Medium Access Control (MAC) and Physical Layer (PHY)
Specific requirements. Amendment 4: Enhancements for Very
High Throughput for Operation in Bands below 6 GHz},
\newblock{IEEE, NewYork, (2013)}


\bibitem{PS}
E. Perahia, R. Stacey,
\newblock{Next Generation Wireless LANs: 802.11n and 802.11ac,}
\newblock{2nd Edition, Cambridge Press, 2013 }

\bibitem{KKL}
E. Khorov, A. Kiryanov, A. Lyakhov,
\newblock{IEEE 802.11ax: How to Build High Efficiency WLANs,}
\newblock{Int. Conf. on Eng. and Telecommunication (2015) 14-19}

\bibitem{AVA}
M. S. Afaqui, E. G. Villegas, E. L. Aguilera,
\newblock{IEEE 802.11ax: Challenges and Requirements for Future
High Efficiency WiFi,}
\newblock{IEEE Wireless Communications 99 (2016) 2-9}

\bibitem{DCC}
D. J. Deng, K. C. Chen, R. S. Cheng,
\newblock{IEEE 802.11ax: Next Generation Wireless Local Area Networks,}
\newblock{10th Int. Conf. on Heterogeneous Networking for Quality, Security and Robustness (QSHINE), (2014) 77-82}

\bibitem{B}
B. Bellalta, 
\newblock{IEEE 802.11ax: High-efficiency WLANs,}
\newblock{IEEE Wireless Communications, 23(1) (2016) 38-46}

\bibitem{SA2}
O. Sharon, Y. Alpert,
\newblock{Scheduling strategies and Throughput optimization for
the Uplink for IEEE 802.11ax and IEEE 802.11ac based networks,}
\newblock{Wireless Sensor Networks,9 (2017) pp. 250-273}

\bibitem{SA3}
O. Sharon, Y. Alpert,
\newblock{Scheduling strategies and throughput optimization for
the Downlink for IEEE 802.11ax and IEEE 802.11ac based networks,}
\newblock{Submitted, Physical Communications}

\bibitem{SA1}
O. Sharon, Y. Alpert,
\newblock{Single User MAC level Throughput comparision: IEEE 802.11ax vs. IEEE
802.11ac,}
\newblock{Wireless Sensor Networks, 9 (2017), pp. 166-177}

\bibitem{BS}
B. Bellalta, K. Kosek-Szott,
\newblock{AP-initiated Multi-User Transmissions in IEEE 802.11ax 
WLANs,}
\newblock{arXiv:1702.05397v1 [cs.NI] (2017)}

\bibitem{SA4}
O. Sharon, Y. Alpert,
\newblock{Optimizing TCP Goodput and Delay in next  generation IEEE
802.11 (ax) devices,}
\newblock{Submitted Ad-Hoc Networks}

\bibitem{KCC}
R. Karmakar, S. Chattopadhyay, S. Chakraborty,
\newblock{Impact of IEEE 802.11n/ac PHY/MAC High Throughput
Enhancement over Transport/Application layer protocols - A Survey,}
\newblock{IEEE Communication surveys and tutorials (2017)}

\bibitem{MKA}
Miorandi, D., Kherani, A. A. and Altman, E. (2006)
\newblock{ A Queueing model for HTTP traffic over IEEE 802.11 WLANs.}
\newblock{Computer Networks, 50, 63-79.}


\bibitem{BCG1}
Bruno, R., Conti, M. and Gregori, E. (2005)
\newblock{Throughput Analysis of UDP and TCP Flows in IEEE 802.11b WLANs: A Simple Model and its Validation.}
\newblock{Workshop on Techniques, Methodologies and Tools for Performance Evaluation of Complex Systems, 2005, 54-63.}


\bibitem{BCG2}
Bruno, R., Conti, M. and Gregori, E. 
\newblock{Throughput Analysis and Measurements in IEEE 802.11 WLANs with
TCP and UDP Traffic Flows.}
\newblock{IEEE Trans. on Mobile Computing, 7, 171-186, (2008)}


\bibitem{KAMG}
Kumar, A., Altman, E., Miorandi, D. and Goyal, M. 
\newblock{New Insights from a Fixed Point Analysis of Single Cell IEEE 802.11 WLANs.}
\newblock{IEEE/ACM Trans. on Networking, 15, 588-601, (2007)}

\bibitem{C}
D. Chen,
{A survey of IEEE 802.11 Protocols: Comparison and Prospective}
\newblock{5th International Conference on Mechatronics, Materials,
Chemistry and Computer Engineering (ICMMCCE) pp. 569-578, (2017)}

\bibitem{OA}
H. A. Omar, K. Abboud,
\newblock{A Survey on High Efficiency Wireless Local Area Networks:
Next Generation WiFi,}
\newblock{IEEE Communications Surveys and Tutorials, (2016)}

\bibitem{NY}
D. Nandal, P. Yadav,
\newblock{A Review on OFDMA and MU-MIMO MAC Protocols for upcoming IEEE
Standard 802.11ax}
\newblock{International Journal on Recent and Innovation Trends in Computing
and Communication, Vol. 4, Issue 8, pp. 108-114 (2016)}


\bibitem{QLYY}
Q. Qu, B. Li, M. Yang, Z. Yan, 
\newblock{An OFDMA based Concurrent Multiuser MAC for Upcoming IEEE 802.11ax,}
\newblock{IEEE Wireless Comm. and Networking Conf. Workshops (WCNCW) pp. 136-141, (2015)}


\bibitem{DLLC}
D.J. Deng, S. Y. Lien, J, Lee, K. C. Chen,
\newblock{On Quality-of-Service Provisioning in IEEE 802.11ax WLANs}
\newblock{IEEE Access, Vol. 4, pp. 6086-6104 (2016)}

\bibitem{AK}
A. Aijaz, P. Kulkarni,
\newblock{Simultaneous Transmit and Receive Operation in Next Generation 
IEEE 802.11 WLANs: A MAC Protocol Design Approach}
\newblock{IEEE Wireless Communications Vol. 24, No. 6, pp. 128-135 (2017)}

\bibitem{PH}
R. Pierre, F. Hoefel,
\newblock{IEEE 802.11ax: On Time Synchronization in Asynchronous OFDM Uplink Multi-User MIMO Physical Layer,}
\newblock{XXXV Simposio Brasileiro De Telecomunicacoes E Processamento De Sinais-Sbr (2017)}

\bibitem{LLYQYZY}
W. Lin, B. Li, M. Yang, Q. Qn, Z. Yan, X. Zuo, B. Yang,
\newblock{Integrated Link-System level Simulation Platform for the
Next Generation WLAN - IEEE 802.11ax,}
\newblock{IEEE Globecom (2016)}


\bibitem{LDC}
J. Lee, D. J. Deng, K. C. Chen,
\newblock{OFDMA-based hybrid channel access for IEEE 802.11ax WLAN,}
\newblock{unpublished.}

\bibitem{KBPSL}
M. Karaca, S. Bastani, B. E. Priyanto, M. Safavi, B. Landfeldt,
\newblock{Resource Management for OFDMA based Next Generation 802.11ax
WLANs,}
\newblock{9th IFIP Wireless and Mobile Networking Conf. (WMNC) (2016)}


\bibitem{JS}
V. Jones, H. Sampath,
\newblock{Emerging technologies for WLAN,}
\newblock{IEEE Commu. Mag., 5 (2015) 141-9 }

\bibitem{RFBBO}
L. Sanabria-Russo, A. Faridi, B. Bellalta, J. Barcelo, M. Oliver,
\newblock{Future evolution of CSMA protocols for the IEEE 802.11
standard}
\newblock{IEEE Int. Conf. on Comm. (ICC) (2013) 1274-9 }

\bibitem{RBFB}
L. Sanabria-Russo, J. Barcelo, A. Faridi, B. Bellalta,
\newblock{WLANs throughput improvement with CSMA/ECA,}
\newblock{IEEE Conf. on Computer Comm. Workshops (INFOCOM WKSHPS)
(2014) 125-6 }

\bibitem{HYSG}
Y. He, R. Yuan, J. Sun, W. Gong,
\newblock{Semi-Random Backoff: Towards resource reservation for
channel access in wireless LANs,}
\newblock{IEEE Int. Conf. on Network Protocols (ICNP) (2009) 21-30 }

\bibitem{KLL}
E. Khorov, V. Loginov, A. Lyakhov,
\newblock{Several EDCA Parameters Sets for Improving Channel Access in
IEEE 802.11ax Networks,}
\newblock{Int. Symposium on Wireless Communication Systems (ISWCS) (2016) 419-423}

\bibitem{SA10}
O. Sharon, Y. Alpert,
\newblock{Coupled IEEE 802.11ac and TCP performance evaluation in various aggregation schemes and Access Categories,}
\newblock{Computer Networks 100 (2016) 141-156}

\bibitem{SA11}
O. Sharon, Y. Alpert,
\newblock{Coupled IEEE 802.11ac and TCP Goodput improvement using Aggregation and Reverse Direction,}
\newblock{Wireless Sensor networks 8(7) (2016) 107-136}

\bibitem{SA12}
O. Sharon, Y. Alpert,
\newblock{Comparison between TCP scheduling strategies in IEEE 802.11ac based Wireless networks,}
\newblock{Ad Hoc Networks 61C (2017) pp. 95-113}

\bibitem{SA}
O. Sharon, Y. Alpert,
\newblock{MAC level Throughput comparison: 802.11ac vs. 802.11n,}
\newblock{Physical Communication 12 (2014) 33-49 }


\bibitem{IEEEn}
\newblock{IEEE Std. 802.11$^{TM}$-2012},
\newblock{Standard for Information Technology - 
Telecommunications and Information Exchange between Systems - Local
and Metropolitan Area Networks - Specific Requirements. Part 11:
Wireless LAN Medium Access Control (MAC) and Physical Layer (PHY)
specifications,}
\newblock{IEEE, NewYork, (2012)}

\bibitem{SA20}
O. Sharon, Y. Alpert,
\newblock{The combination of aggregation, ARQ, QoS guarantee and
mapping of Application flows in very High Throughput 802.11ac networks,}
\newblock{Physical Communication  17 (2015) 15-36 }

\bibitem{IEEEber}
IEEE P802.11 Wireless LANs, 11ax Evaluation Methodology, Appendix 3 -
RBIR and AWGN PER Tables (2015) .

% \bibitem{ns2}
% \newblock{Network Simulator 2 (NS-2). Available:}
% \newblock{http://www.isi.edu/nsnam/ns.}

% \bibitem{M}
% E. Modiano,
% \newblock{An adaptive alogorithm for optimizing the packet size used in wireless ARQ protocols,}
% \newblock{Wirel. Netw. 5 (1999) 279-286.}

% \bibitem{YZZZ}
% F. Yang, Q. Zhang, W. Zhu, Y.Q. Zhang,
% \newblock{An efficient transport scheme for multimedia over wireless internet,}
% \newblock{IEEE Int. Conf. on 3G wireless and beyond (3Gwireless), 2001}

% \bibitem{RS}
% K. Ramin, K. Sabmatian,
% \newblock{Evaluation of packet error rate in wireless networks,}
% \newblock{7th IEEE/ACM Int. Sym. on Mobility, Analysis and Simulation of Wireless Mobile Systems (MSWIM), 2004}

% \bibitem{GAGM}
% D. Gomez, R. Aguero, M. Garcia-Arranz, L. Munoz,
% \newblock{Replication of the Bursty Behavior of Indoor WLAN Channels,}
% \newblock{Proc. of the 6th Int. ICST Conf. on Simulation Tools and Techniques, (2013) pp. 219-226}


% \bibitem{WIFI}
% \newblock{The Wi-Fi Alliance, http://www.wi-fi.org }

% \bibitem{L1}
% J. Lemmon,
% \newblock{Wireless link statistical bit error rate model}, 
% \newblock{Technical Report 02-934, U.S. Dept. of Commerce,
% June, (2002)}

% \bibitem{CBV}
% P. Chatzimisios, A. C. Boucouvalas, V. vitas,
% \newblock{IEEE 802.11 Packet Delay - A Finite Retry Limit Analysis,}
% \newblock{in Proc. IEEE Globecom, 2 (2003), 950-954 }


% \bibitem{CBV}
% P. Chatzimisios, A. C. Boucouvalas, V. vitas,
% \newblock{IEEE 802.11 Packet Delay - A Finite Retry Limit Analysis,}
% \newblock{in Proc. IEEE Globecom, 2 (2003), 950-954 }

% \bibitem{SA5}
% O. Sharon, Y. Alpert, 
% \newblock{Comparison between TCP scheduling strategies in IEEE 802.11ac based Wireless networks,}
% \newblock{AD HOC networks 61C (2017) pp. 95-113}

\end{thebibliography}

%%%%%%%%%%%%%%%%%%%%%%%%%%%%%%%%%%%%%%%%%%%%%%%%%%%%%%%%%%%%%%%%

\end{document}